# Dynamics of non-thermal states in optimally-doped $Bi_2Sr_2Ca_{0.92}Y_{0.08}Cu_2O_{8+\delta}$ revealed by mid-infrared three-pulse spectroscopy

Angela Montanaro[1,2,3], Enrico Maria Rigoni[1,2], Francesca Giusti[1,2], Luisa Barba[4], Giuseppe Chita[4], Filippo Glerean[5], Giacomo Jarc[1,2], Shahla Y. Mathengattil[1,2], Fabio Boschini[6,7], Hiroshi Eisaki[8], Martin Greven[9], Andrea Damascelli[7,10], Claudio Giannetti[11,12], Dragan Mihailovic[13], Viktor Kabanov[13], and Daniele Fausti[1,2,3*]

[1]*Department of Physics, Università degli Studi di Trieste, 34127 Trieste, Italy*
[2]*Elettra Sincrotrone Trieste S.C.p.A., 34149 Basovizza Trieste, Italy*
[3]*Department of Physics, University of Erlangen-Nürnberg, 91058 Erlangen, Germany*
[4]*Institute of Crystallography, CNR, Elettra Sincrotrone Trieste S.C.p.A., 34149 Basovizza Trieste, Italy*
[5]*Department of Physics, Harvard University, Cambridge, Massachusetts 02138, USA*
[6]*Centre Énergie Matériaux Télécommunications, Institut National de la Recherche Scientifique, Varennes, Québec, Canada J3X1S2*
[7]*Quantum Matter Institute, University of British Columbia, Vancouver, BC, Canada V6T 1Z4*
[8]*Nanoelectronics Research Institute, National Institute of Advanced Industrial Science and Technology, Tsukuba, Ibaraki 305-8568, Japan*
[9]*School of Physics and Astronomy, University of Minnesota, Minneapolis, Minnesota 55455, USA*
[10]*Department of Physics & Astronomy, University of British Columbia, Vancouver, BC, Canada V6T 1Z1*
[11]*Department of Mathematics and Physics, Università Cattolica, I-25121 Brescia, Italy*
[12]*Interdisciplinary Laboratories for Advanced Materials Physics (I-LAMP), Università Cattolica, I-25121 Brescia, Italy*
[13]*Jožef Stefan Institute, Jamova 39, 1000 Ljubljana, Slovenia*

*\*Correspondence: daniele.fausti@elettra.eu*

## ABSTRACT

**In the cuprates, the opening of a *d*-wave superconducting (SC) gap is accompanied by a redistribution of spectral weight at energies two orders of magnitude larger than this gap. This indicates the importance to the pairing mechanism of on-site electronic excitations, such as orbital transitions or charge transfer excitations. Here, we resort to a three-pulse pump-probe scheme to study the broadband non-equilibrium dielectric function in optimally-doped $Bi_2Sr_2Ca_{0.92}Y_{0.08}Cu_2O_{8+\delta}$ and we identify two distinct dynamical responses: i) a blueshift of the central energy of an interband excitation peaked at 2 eV and ii) a change in spectral weight in the same energy range. Photoexcitation with near-IR and mid-IR pulses, with photon energies respectively above and below the SC gap, reveals that the transient changes in the central energy are not modified by the onset of superconductivity and do not depend on the pump photon energy. Conversely, the spectral weight dynamics strongly depends on the pump photon energy and has a discontinuity at the critical temperature. The picture that emerges is that, while high-energy pulses excite quasiparticles in both nodal and thermally inaccessible antinodal states, photoexcitation by low-energy pulses mostly accelerates the condensate and creates excitations predominantly at the nodes of the SC gap. These results, rationalized by kinetic equations for *d*-wave superconducting gaps, indicate that dynamical control of the momentum-dependent distribution of non-thermal quasiparticles may be achieved by the selective tuning of the photoexcitation.**

## I. INTRODUCTION

The interplay between the high- and low-energy electrodynamics is one of the unconventional features of cuprate superconductors. While in standard BCS superconductors the opening of a superconducting (SC) gap ($\Delta_{SC}$) triggers a rearrangement of the quasiparticle excitation spectrum only in an energy range that is comparable with $2\Delta_{SC}$ (a few tens of meV), the condensate formation in the cuprates is accompanied by a redistribution of spectral weight at energies that are two orders of magnitude larger than $2\Delta_{SC}$ [1-6]. This anomalous and still much-debated behaviour hints at a direct involvement of the high-energy electronic excitations in the pairing mechanism of high-$T_C$ superconductivity.

A further complication arises from the symmetry of the SC gap. Whereas standard BCS superconductors exhibit a fully symmetric *s*-wave gap that naturally emerges from phonon-mediated pairing, in the cuprates the gap has *d*-wave symmetry [7-10]. Whereas at the antinode electronic transitions are permitted only at frequencies larger than the SC gap, electronic excitations at all frequencies are allowed at the nodes, where the gap amplitude vanishes.

Understanding how the momentum distribution of the electronic excitations affects the *d*-wave SC gap is pivotal to advance the field and, arguably, a way to distinguish between the many microscopic models proposed for the cuprate superconductors [11-13]. This has stimulated a tremendous development of momentum-resolved experimental techniques, both equilibrium and their time-resolved counterparts [14-16].

When the cuprates are photoexcited by ultrashort laser pulses, excess quasiparticles (QPs) are injected into the system and perturb the initial equilibrium distribution [17-27]. While ARPES studies indicate that high-photon energy pulses (>1.5 eV) excite a QP population uniformly distributed across the Fermi surface [26, 28], the effect of the photoexcitation by low-photon energy pulses (i.e., smaller than the SC gap) has been less investigated [29]. Evidence for transient SC phases triggered by long-wavelength electric fields has been reported [30-35], suggesting that excitation by pulses with photon energy smaller than the SC gap may lead to different transient electronic states.

To disentangle the effects of high- and low-photon energy excitations, here we employ a three-pulse scheme that combines a near-infrared pump (near-IR, $h\nu_{NIR} \gg 2\Delta_{SC}$), a mid-infrared pump (mid-IR, $h\nu_{MIR} \leq 2\Delta_{SC}$), and a white-light probe ($1.5 < h\nu < 2$ eV). The rationale of our approach is to first use one pump to drive the system into a non-thermal state, and then to measure the subsequent QP dynamics with the second pump-probe sequence. The broadband detection window enables the measurement of the transient high-energy dielectric function of the sample, and thus allows us to directly probe the involvement of the high-energy electronic excitations in that spectral range.

We find that the transient response in optimally-doped $Bi_2Sr_2Ca_{0.92}Y_{0.08}Cu_2O_{8+\delta}$ (Y-Bi2212) below $T_C$ can be fully accounted for by a blueshift and an impulsive change of spectral weight of an interband excitation at 2 eV. Importantly, while the blueshift dynamics is qualitatively the same regardless of the choice of the pump and the sample temperature, the dynamics of the spectral weight (*SW*) is sensitive to both the pump photon energy and the pump polarization and displays a clear signature of the onset of the SC state.

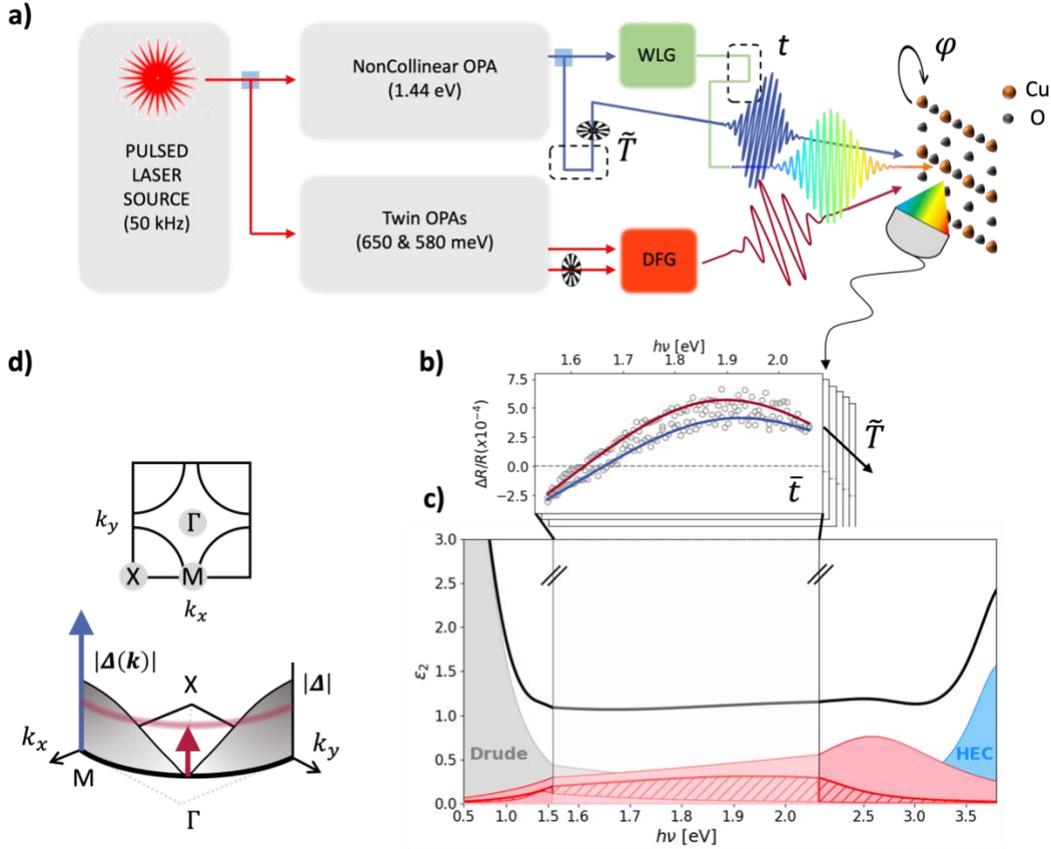

**Figure 1: Time-dependent broadband measurements of dielectric function upon photoexcitation above and below the superconducting gap. a)** Sketch of the three-pulse experimental setup. "WLG" stands for White-Light Generation, "DFG" for Difference-Frequency Generation. "$\tilde{T}$" indicates the delay between the two pumps, "$t$" the delay between the near-IR pump and the supercontinuum probe. The in-plane sample orientation ($\varphi$) can be adjusted by a piezoelectric rotator mounted. **b)** Example of transient reflectivity spectra (grey circles) and corresponding differential fits to the data (solid lines) at fixed $\bar{t}$ for the near-IR (blue) and mid-IR (red) pumps, simultaneously acquired. This is representative of the full $\tilde{T}$-dynamics. **c)** Imaginary part of the dielectric function obtained through a Kramers-Kronig constrained Drude-Lorentz fit [52] of the dielectric function of Y-Bi2212 at T=20 K (data published in Ref. [46]). The grey-shaded area accounts for the Drude peak and includes the low-energy contributions. The blue-shaded area indicates the high-energy contributions (HEC). The red peaks correspond to the oscillators that overlap with our probe window. The red pattern indicates the oscillator at $\omega_0 = 2$ eV that is the focus of our studies. **d)** First Brillouin zone and absolute value of the superconducting gap in the reciprocal space for a $d$-wave superconductor. The blue and the red arrows represent the photoexcitations above and below the gap, respectively.

We use the *SW* dynamics as an observable to study the response of the sample to subsequent excitations by the near-IR and the mid-IR pumps. We observe that the double-pump response depends on the order in which the two pump impinge on the sample. When the system is first photoexcited by sub-gap pulses, the *SW* dynamics triggered by the near-IR pump is not affected by the presence of the previous photoexcitation. Conversely, photoexcitation by above-gap pulses strongly modifies the dynamics later initiated by the mid-IR pump, which becomes similar to the one triggered by high-energetic pulses. Overall, our measurements can be rationalized within a theoretical framework based on the kinetic equations for superconductors [36-38], which can discriminate between the effects of low- and high-frequency photoexcitations on the momentum-resolved distribution function of non-equilibrium QPs.

The manuscript is organized as follows. The experimental technique combining mid-IR and near-IR pumps with a broadband probe is presented in Section II. Section III reports the main experimental evidence and the discussion of our results. The visible spectral response of Y-Bi2212 is discussed in IIIA, the dynamical response to single photoexcitation at different pump wavelength in the SC phase in IIIB, and in the normal and pseudogap phase in

IIIC. The temperature dependence of the response to photoexcitation with single pump is discussed in IIID, while the double-pump measurements and kinetic model are reported and discussed in Section IIIE and IIIF, respectively.

## II. EXPERIMENTAL METHODS

We carried out time-domain optical measurements on a freshly cleaved sample of optimally-doped Y-Bi2212 at different temperatures. The sample displays a SC phase below $T_C$ = 96 K [39], and a pseudogap (PG) phase between $T_C$ and T* ≅ 135 K; above T*, the system behaves as a "strange metal". The absolute value of the $d$-wave gap in reciprocal space is sketched in Fig. 1d. The antinodal gap amplitude is $2\Delta_{SC}$ ~ 75 meV [40].

The three-pulse optical setup is sketched in Fig. 1a. The sample is simultaneously photoexcited by a near-IR pump (1.44 eV ≫ $2\Delta_{SC}$) generated by a non-collinear parametric amplifier system and a mid-IR pump (70 meV ≲ $2\Delta_{SC}$) obtained through difference frequency generation of two near-IR pulses coming from a twin optical parametric amplifier system. The transient broadband reflectivity is probed by a white-light supercontinuum (1.5-2 eV) obtained through self-phase modulation in a sapphire crystal. The three pulses propagate along the $c$-axis (orthogonally to the Cu-O plane). The pump beams are co-polarized, while the probe polarization is orthogonal to the pumps; the polarization of the impinging beams with respect to the $CuO_2$ plaquette can be adjusted by rotating the sample with a piezoelectric rotator in the vacuum chamber. The sample has been oriented through X-ray diffraction measurements at the beamline XRD1 at Elettra Sincrotrone Trieste. More details on the experimental setup can be found in Ref. [41].

The two optical choppers placed along the pumps' optical paths run such that one is twice as fast as the other. A proper sorting of the probe pulses according to the choppers' rotational frequencies allows to single out i) the response to only the near-IR pump; ii) the response to only the mid-IR pump; iii) the joint response to both the photoexcitations. As illustrated in Fig. 1b, for each given delay $\bar{t}$ of the probe beam, the broadband transient change in reflectivity induced by the pumps is measured. By tuning the time delay $\tilde{T}$ between the pump pulses, the order of arrival of the pumps can be swapped. This allows us to determine, within the same measurement, how the photoexcitation by the first impinging pump modifies the dynamics measured by the second pump-probe sequence.

## III. RESULTS AND DISCUSSION

### A. EQUILIBRIUM OPTICAL PROPERTIES OF Y-Bi2212

The equilibrium properties of the cuprates are mainly determined by the Cu 3$d$ and the O 2$p$ orbitals within the $CuO_2$ plaquette. In the insulating parent compounds, the strong on-site Coulomb repulsion splits the Cu 3$d$ band into a lower (LHB) and an upper (UHB) Hubbard band. The O 2$p$ band, which is fully occupied, falls inside this gap, ~2 eV below the UHB. The transition from the O 2$p$ band to the UHB, which is the lowest-energy one in the system, represents the charge-transfer (CT) of a Cu $3d_{x^2-y^2}$ hole to its neighbouring O $2p_{x,y}$ orbitals and marks the lower bound of the electron-hole continuum. Upon hole doping, the charge-transfer edge ($\Delta_{CT}$) shifts at higher energies and the CT gap is filled with states [42,43]. The assignment of the excitations within the energy window spanning from 1.5 to 2 eV is still controversial. Based on dynamical mean field calculations, it has been proposed that the interband excitations are mostly related to many-body hybridized Cu-O states, corresponding to the Zhang-Rice singlet states [44-46]. On the other hand, $dd$ orbital transitions are also expected to give a contribution in the range 1.5-2 eV. Evidence for an excitation localized at these energies has been reported by Raman scattering experiments in a large number of undoped cuprate superconductors and attributed to the Raman-activated hole-transition from the in-plane $d_{x^2-y^2}$ to the $d_{xy}$ orbital with $A_{2g}$ symmetry [47]. Furthermore, resonant inelastic x-ray scattering measurements showed that the $dd$ orbital transitions dominate the excitation spectrum at these energies also in doped SC cuprates [48,49]. The recent observation of a modification of the *dd* spectrum upon entering the SC phase further emphasizes the role that these excitations may play in the pairing [50,51].

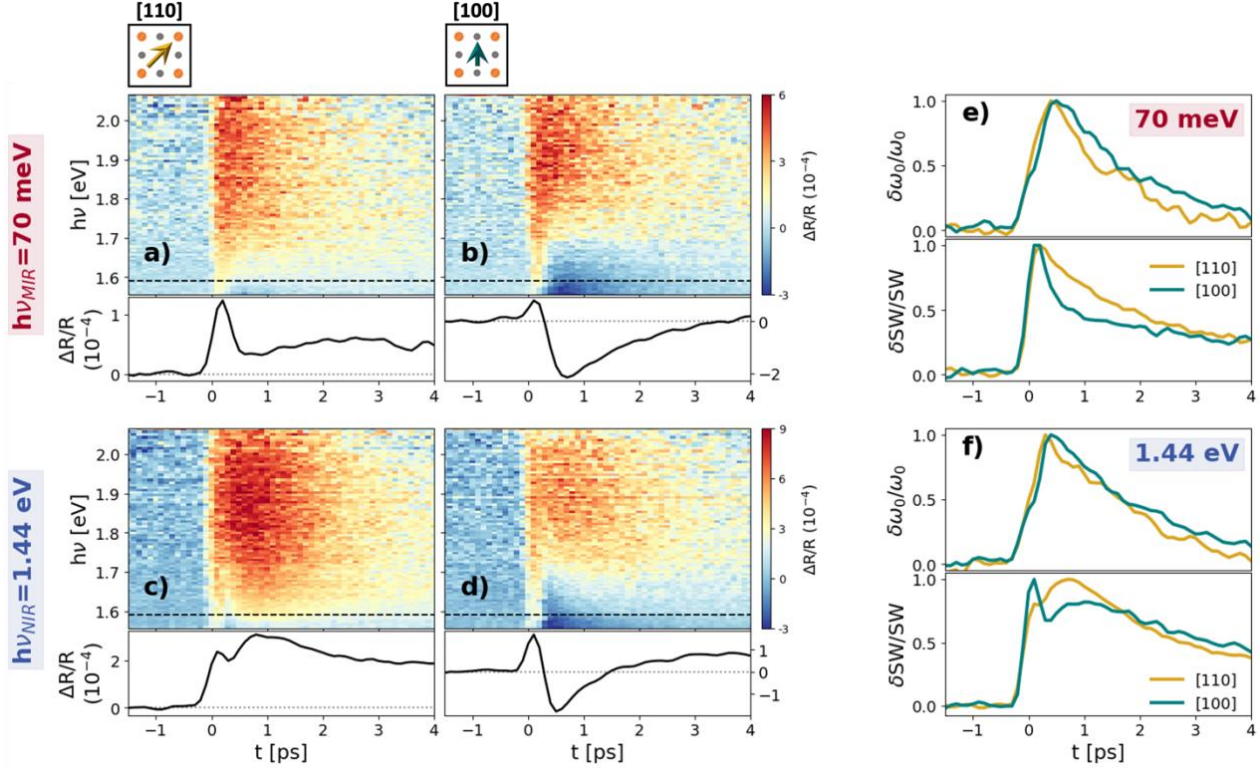

**Figure 2: Polarization-dependent pump-probe measurements for below- and above-gap photoexcitation in the superconducting phase. a-b)** Broadband transient reflectivity upon a photoexcitation by mid-IR pulses (70 meV) with a pump pulse polarized along [110] and [100], respectively. The black line in the lower panels is a cut at 1.59 eV. The fluence of the mid-IR pump was $\phi_{MIR}=28$ μJ/cm$^2$ and the sample temperature was T=74 K. **c-d)** Same as a-b), but for a high-energy near-IR photoexcitation (1.44 eV) and fluence $\phi_{NIR}=19$ μJ/cm$^2$. **e)** Results of differential fits of the measurements in a) and b), as described in the text. For both polarizations of the pump ([110] in gold and [100] in teal), the curves in the top panel show the normalized transient change in the central frequency of the oscillator at $\omega_0= 2$ eV, which is primarily responsible for the non-equilibrium response in the probed spectral region, as shown in Fig. 1c. In the bottom panel, the curves represent the normalized transient change in the integrated spectral weight ($SW \propto \int_0^{5eV} d\omega\, \omega\, \varepsilon_2(\omega)$,) for the two pump polarizations. **f)** Same as e), but for the near-IR photoexcitation.

In Fig. 1c, we plot the imaginary part of the dielectric function of an optimally-doped Y-Bi2212 sample measured at T=20 K by spectroscopic ellipsometry (adapted from Ref. [46]). While the low-energy side ($h\nu < 1.25$ eV) of the dielectric function is dominated by the tail of the Drude response of the itinerant carriers and by lower-energy mid-IR contributions, the high-energy side is reminiscent of a CT-like absorption edge ($h\nu > 2.5$ eV). The interband transitions give rise to a structured absorption in the region extending from 1.25 to 2 eV. As indicated by the grey-shaded area, our broadband probe overlaps with this window, directly enabling the measurement of the impulsive modification of these states.

In order to quantitatively estimate the pump-induced changes in the electronic excitations, we fitted the static dielectric function of the material using a Drude-Lorentz model (see Supplementary Note 1) consisting of a Drude peak ($\varepsilon_D$) and a sum of Lorentzian oscillators to account for the electronic transitions ($\sum_i \varepsilon_L^i$). In agreement with Ref. [46], we found that three oscillators are sufficient to fit the dielectric function in the energy range 1.25-2 eV and they correspond to the three red Lorentzian peaks in Fig. 1c. We highlight with a red pattern the transition peaked at 2 eV that will be the focus of our time-domain studies. It is worth noting that the transitions extrapolated from the fit could also be made of several unresolved peaks, as the complexity of the optical spectra of the cuprates might suggest [53]. The Lorentzian peaks in Fig. 1c should be therefore considered as an effective description rather than a physical microscopic modelling of the complex optical response of the sample. The discussion in the present work does not aim to clarify the nature of these excitations and will be limited to considering to what extent the high-energy physics is involved in the pairing.

## B. SELECTIVE PHOTOEXCITATION IN THE SUPERCONDUCTING PHASE

The study of the dynamics in a tailored non-thermal state relies on the knowledge of the QP dynamics in the system that is initially at equilibrium. In this section, we focus on the characterization of the sample response to a single excitation with photon energy either below or above the SC gap.

In Fig. 2a,c, we compare the broadband transient reflectivity upon photoexcitation in the SC phase by the mid-IR pump and the near-IR pump, respectively. The polarization of both pumps was parallel to [110], as illustrated in the top inset. The measurements were performed at low pumping fluences ($\phi < 30$ μJ/cm$^2$) to avoid the complete vaporization of the condensate [19]. To guarantee a meaningful comparison between the signal at high- and low-pump photon energy, we tuned the intensity of the pumps to have similar ΔR/R amplitude at fixed probe energy. Although the transient change in reflectivity is positive over the entire probe window (in agreement with the one measured in ref. [46] with a 1.55 eV pump pulse), the dynamics of the signal displays a dependence on the photon energy of the pump. This difference is more pronounced at low probe energies, where the black cuts (hv$_{pr}$ = 1.59 eV) in the lower panels of Fig. 1a,c are selected.

Motivated by the distinctive gap anisotropy of the sample and its anisotropic response to long-wavelength pulse excitation that we have recently reported [35, 54], we investigated how the broadband response changes according to the polarization of the pump beams. In Fig. 2b,d, we show the time- and spectrally-resolved transient reflectivity for the pump polarizations parallel to the direction of the Cu-O bond ([100]). We observe that, in both maps, the transient reflectivity changes sign at approximately 1.7 eV and displays a negative dynamics below this threshold (black cuts in the lower panels of Fig. 2b,d). We stress that reducing the pump and the probe fluences does not qualitatively affect the spectral dependence of the response in this polarization configuration (Fig. S1-S3), confirming that the measurements are performed in the linear regime.

To understand the origin of the observed structured reflectivity, we performed a differential fit of the maps based on the equilibrium dielectric function ($\varepsilon_{eq}(\omega)$) discussed in Section IIIA. The key idea of the analysis is to modify the minimum number of parameters that model $\varepsilon_{eq}(\omega)$ in order to fit, at every time delay, the measured transient broadband reflectivity (see Supplementary Note 2 for details on the fitting procedure). The analysis revealed that a transient change of only the Lorentzian oscillator peaked at 2 eV is sufficient to reproduce our data. In particular, we found that the photoexcitation results in a blueshift of the central frequency ($\omega_0$) of this oscillator and in a change of the integrated spectral weight (*SW*), formally defined as $SW \propto \int_0^{5eV} d\omega\, \omega\, \varepsilon_2(\omega)$, where we performed the integration over the whole energy axis of the fitted $\varepsilon_{eq}(\omega)$ (Supplementary Note 1). We stress that the dynamics of $\omega_0$ and *SW* are uncorrelated and thus these two parameters constitute a good pair of candidates to consistently describe the measurements (Supplementary Note 2).

We show in Fig. 2e,f the results of the differential fits for photoexcitation with pulse photon energy below and above $2\Delta_{SC}$, respectively. We plot the relative change $\delta\omega_0/\omega_0$ and $\delta SW/SW$ as a function of the pump-probe delay for both pump polarizations in each panel and, to facilitate the comparison between the curves, we normalize the dynamics by the maximum positive value. Neither the pump photon energy nor the pump polarization have any influence on the dynamics of the blueshift: the dynamics in all the four combinations (curves in the top panels of Fig. 2e,f) can be well described by a single-exponential with a decay time equal to ∼1.5 ps.

The dynamics of the integrated *SW* is instead more peculiar. By comparing the curves in the bottom panels of Fig. 2e and 2f, we observe a different dynamics depending on the pump used: when the sample is photoexcited by the sub-gap pump (bottom panel in Fig. 2e), the *SW* dynamics displays a single-exponential decay, while the dynamics initiated by the above-gap pump (bottom panel in Fig. 2f) is characterized by a fast-decaying component and an additional slower decay.

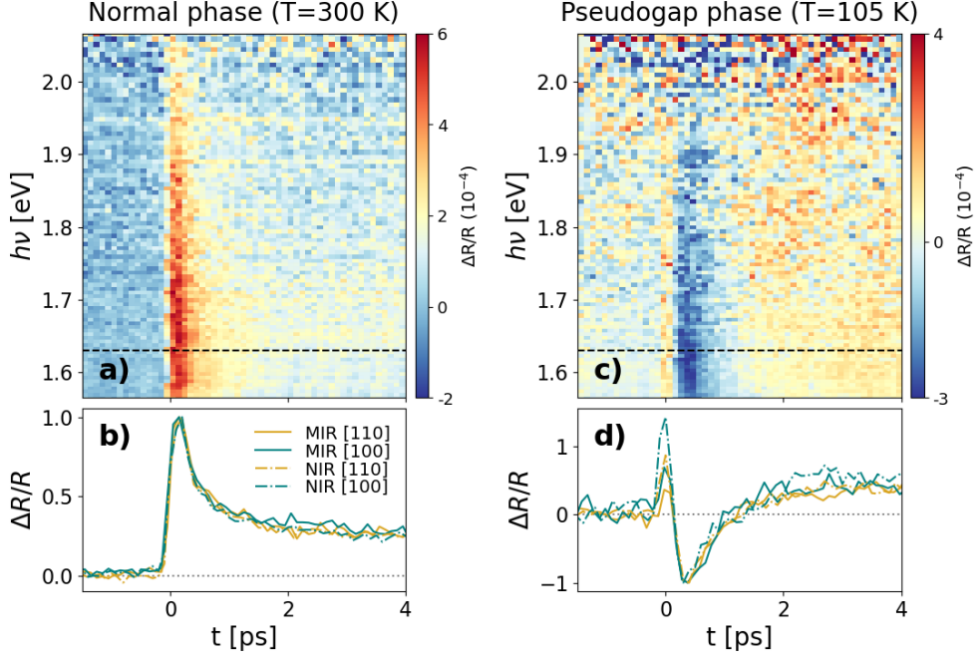

**Figure 3: Sub-gap pump-probe measurements in normal and pseudogap phases. a)** Transient reflectivity map at T=300 K upon photoexcitation by the mid-IR pump polarized along [110] ($\phi_{MIR}$=28 µJ/cm$^2$). **b)** Normalized cuts at 1.63 eV for the four configurations of pump photon energy (mid-IR or near-IR) and pump polarization ([110] or [100]). **c-d)** Same as a-b), but in the pseudogap phase at T=105 K.

It is interesting to note that the behaviour in the bottom panel of Fig. 2f is reminiscent of the dynamics of the transient single-colour reflectivity (at 1.55 eV) measured in other works upon photoexcitation by above-gap pumps [19,55]. In particular, the fast peak is commonly ascribed to the incipient melting of the SC phase due to the sudden photo-injection of hot QPs that non-thermally lead to the filling of the gap [56,57]. The emergence of the second peak, which is farther delayed in time the more intense the photoexcitation is, is instead usually associated with the recovery dynamics of the condensate. In Bi2212, this has been observed to set in within a few ps [58]. However, this peculiar transient reflectivity response is not a generic sample's response and ultimately depends on the experimental parameters, such as probe photon energy and polarization [55,59]. Here, by fitting the transient dielectric function and thereby overcoming the limitations intrinsic to single-colour pump-probe measurements, we show that this dynamics is directly mapped onto the variation of the integrated spectral weight (probe-energy independent quantity), and we thus confirm its general validity.

Interestingly, the absence of such dynamics in the mid-IR-induced *SW* change (bottom panel of Fig. 2e) may be an indication that the recombination is different in this case. It is generally accepted that excitation by high photon energy pulses uniformly targets the Fermi surface, leading to an accumulation of photo-injected QPs both at the nodes and at the antinodes of the SC gap [26,28]. However, as the double-pump measurements will show, photoexcitation by sub-gap pulses might not inject enough energy to excite QPs at the antinodes, leading to an anisotropic QP population mostly localized in the nodal regions of the Brillouin zone. We note that increasing the fluence of the mid-IR pump does not alter the dynamics of the photo-induced change in *SW* (Fig. S4), which still lacks the recovery component.

Finally, the *SW* dynamics is also affected by the pump polarization. For both pump photon energies, the dynamics is faster when the pump is polarized along [100], which corresponds to the antinodal direction. This further suggests that, in this condition, the photo-induced quench of the SC gap is more efficient [35,54].

## C. NORMAL AND PSEUDOGAP PHASE

The peculiar dependence of the broadband reflectivity on the photon energy and polarization of the pump pulse is a property of the SC phase. In Fig. 3a,c we show the time- and spectrally-resolved reflectivity measured in the normal (T=300 K) and the PG (T=105 K) phase upon photoexcitation by the mid-IR pump polarized along [110]. The normal phase is characterized by a positive response that decays faster than the one in the SC phase. This is consistent with the literature [60] and generally attributed to the suppression of recombination processes due to the opening of the gap below $T_C$. In Fig. 3b, we plot the normalized time trace at $h\nu_{pr}$ = 1.63 eV measured for both the pump polarizations and photon energies. The temporal traces at the same probe energy at T=105 K (Fig. 3d) indicate that the transient reflectivity in the PG phase is qualitatively the same in all the configurations. After an initial positive peak, the reflectivity is negative; after ∼2 ps, a positive signal is restored. The lack of any pump dependence of the signal further suggests that the anisotropies emerging in the SC phase (Fig. 2) are associated with the symmetry of the SC gap.

Similar to the differential analysis carried out in the SC phase, we fitted the maps in Fig. 3a,c based on the dielectric function measured at T=300 K and T=100 K in Ref. [46]. Our analysis shows that, if the interband oscillators are arbitrarily frozen, the transient reflectivity above $T_C$ can be fitted by modifying only the effective scattering rate of the Drude peak ($\Gamma_D$) in the dielectric function (Supplementary Note 3), in agreement with other works [25,46,61].

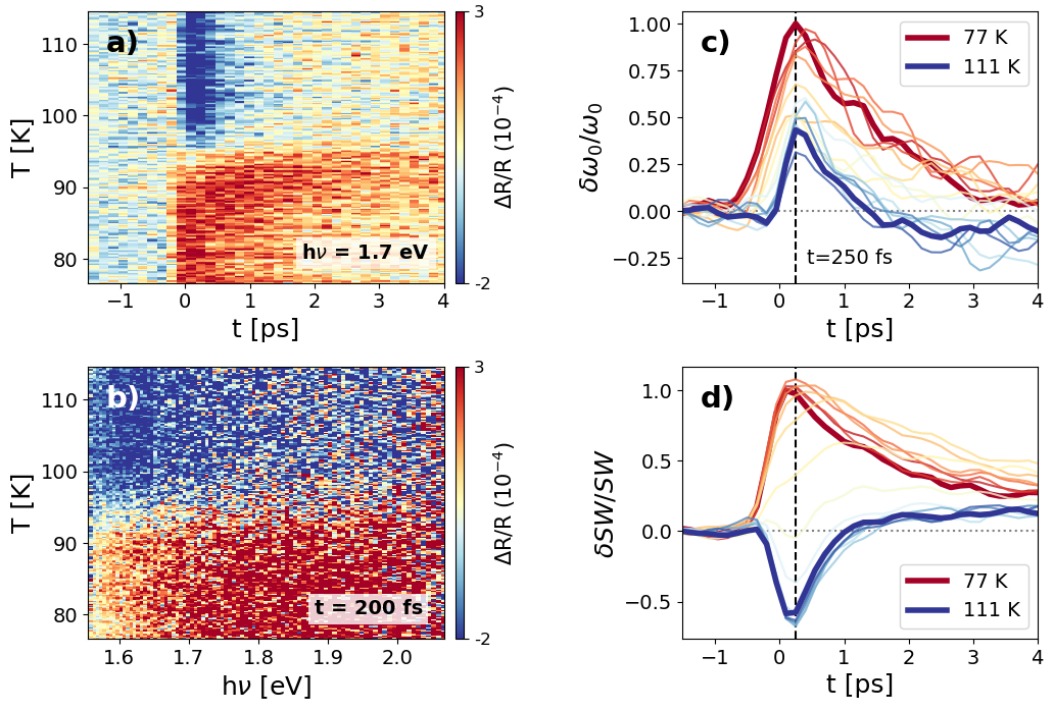

**Figure 4: Temperature-dependent dynamics of the interband excitation. a)** Transient reflectivity as a function of temperature and time delay at fixed probe energy (1.7 eV) upon photoexcitation by the mid-IR pump polarized along [110] ($\phi_{MIR}$=28 µJ/cm$^2$). **b)** Same as a) but as function of temperature and probe energy at fixed time delay (200 fs). **c)** Time-dependent normalized transient change in the central frequency of the $\omega_0$=2 eV oscillator for selected temperatures measured in a). **d)** Time-dependent normalized transient change in spectral weight for selected temperatures in a).

This outcome, however, poses the delicate question of how to deal with the superconducting-to-pseudogap transition. When the PG phase is approached from higher temperatures, one effective parameter ($\Gamma_D$) would seem sufficient to track the transition from the normal phase. This is not the case at $T_C$, where the differential fits instead suggest a discontinuity in our model. By performing temperature-dependent measurements across the SC transition, we will show in the next section that a transient change of both the central frequency and the spectral weight of the 2 eV oscillator can equivalently account for the signal in the PG phase. However, the bandwidth of our measurements is not sufficient to conclusively address this question.

### D. NON-EQUILIBRIUM RESPONSE ACROSS THE SUPERCONDUCTING PHASE TRANSITION

In Fig. 4a, we plot the relative reflectivity variation as a function of the pump-probe delay (horizontal axis) and the sample temperature (vertical axis) across the SC transition. As an example, we plot here only the response to the sub-gap pump polarized along [110] and refer to Fig. S5,S6 for a complete characterization of the temperature-dependent response under other experimental conditions. At each temperature, the broadband response of the sample is measured by the white-light probe (Fig. 4b) and we select one specific probe energy (1.7 eV) to plot the color-coded map in Fig. 4a. A clear discontinuity near $T_C$ is present, where the signal changes sign. However, this is not a universal feature of the optical response: by changing either the pump or the probe photon energy and the pump polarization, the signal qualitatively changes (Fig. S5,S6) and the SC transition cannot be readily inferred from the optical measurements.

We fitted the reflectivity maps using the model described in Section IIIA (Fig. S7). The results of the analysis are plotted in Fig. 4c,d, where we show the dynamics of $\omega_0$ and $SW$, respectively, of the 2 eV transition at different temperatures, from the SC phase (red curves) to the PG one (blue curves). The dynamics of the central frequency of the oscillator does not display a strong dependence on the temperature and remains positive also above $T_C$. This is in contrast to the $SW$ dynamics, which changes sign when $T_C$ is approached.

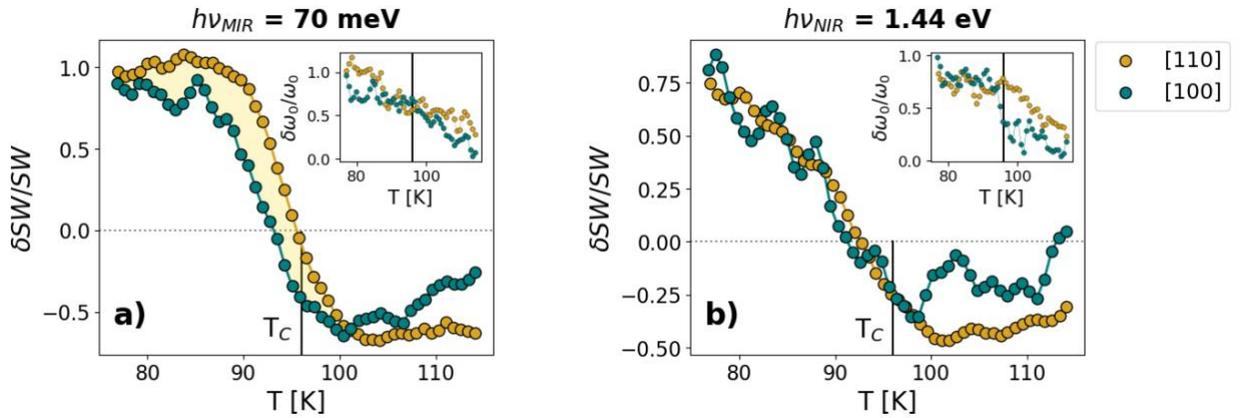

**Figure 5: Pump-induced change of the interband excitation across the superconducting transition. a-b)** Normalized transient change of the spectral weight of the 2 eV oscillator at t=250 fs as function of temperature induced by the below- and the above-gap excitation, respectively, polarized along [110] (gold) and [100] (teal). The black vertical lines indicate the critical temperature. The insets show the temperature-dependent transient change of the central frequency of the oscillator at t=250 fs.

In Fig. 5, we plot the central frequency and the integrated spectral weight as a function of temperature associated with the maximum pump-induced change (t=250 fs). From the main panels of Fig. 5a and b, it is evident that the impulsive change of $SW$ of the 2 eV interband oscillator is a good indicator of the SC transition. In contrast, the insets of Fig. 5 show the temperature dependence of the effective parameter $\omega_0$, which reveals no transitions at $T_C$ regardless of the pump photon energy and polarization.

This analysis emphasizes once again how the condensate formation is linked to the local high-energy excitations in the material [1,2,62,63]. In fact, it is known that in equilibrium a transfer of spectral weight from the high-energy spectral region to the zero-frequency δ-function of the condensate occurs at $T_C$ [64,65]. Following the same interpretation, the pump-induced spectral weight change we observe below $T_C$ should be interpreted as a direct consequence of the quench of the SC pair density, which in our measurements does not appear to be directly related to the shift of the central frequency of the 2 eV oscillator.

In this respect, it is important to note that the dynamics of the central frequency of the oscillator at 2 eV and the integrated spectral weight in the visible range are indicative of two distinct processes, the first one independent of the presence of the condensate and the second one appearing only once three-dimensional superconductivity is observed. We conjecture that the light-driven shift of the oscillator is visible when the spectrum of electronic excitations is locally gapped, while the light-driven spectral weight dynamics is associated with the inductive response of the system which is related instead to the number of electrons in the condensate.

Interestingly, the excitation by above- and below-gap pumps yields different results. While high-photon energy pulses (regardless of their polarization) quench the condensate at temperatures lower than the equilibrium $T_C$ (Fig. 5b), sub-gap pulses have an anisotropic effect (Fig. 5a). When the mid-IR pump is polarized along [100], the quench is similar to the one induced by high-photon energy pulses. However, when the mid-IR polarization is parallel to [110], the discontinuity in the optical response associated to the SC transition occurs at 2-3 K higher temperatures (gold markers in Fig. 5a).

The scenario that emerges is consistent with the previous observation [35] of a superconducting-like response in single-colour reflectivity measurements enhanced by mid-IR pulses polarized along [110]. In this regard, the employment of a broadband probe and thus the use of the integrated spectral weight as an observable (rather than the reflectivity at only one probe wavelength) enables us to provide a more general result. Importantly, our measurements highlight that the anisotropy of the sub-gap pulses arises from the anisotropic response of the high-energy electronic excitation to the perturbation, which in turn affects the dynamics of the condensate.

### E. OPTICAL RESPONSE OF NON-THERMAL STATES

The difference found between the excitation by above- and sub-gap pulses suggests that the transient non-thermal states photo-induced by these two perturbations are different. In particular, the single-pump measurements showed that i) the dynamics of the spectral weight of the 2 eV oscillator photo-induced by the near-IR and mid-IR pump is different in the SC phase and that ii) its temperature-dependence marks the onset of superconductivity and suggest that high-photon energy pulses dynamically quench the condensate more efficiently.

In order to understand the origin of these differences, we study here the dynamical response of a sample previously prepared in a non-thermal state by the photoexcitation with an initial impinging pulse. The rationale of this approach is that, depending on the photon energy of the first pump, the resulting non-thermal QP distribution across the Brillouin zone is expected to be different. A second pump-probe sequence is then employed to measure the broadband transient response of the photo-induced non-thermal state. We delay the second pump-probe sequence by 1 ps with respect to the first pump: this choice guarantees that the initially excited QPs are not completely relaxed before the arrival of the second pump [58] and, at the same time, avoids possible coherent artefacts arising from a temporal overlap of the two pumps. We plot in Fig. 6 the results of these measurements. For clarity, we show only the dynamics of the *SW* obtained through the differential fits, as we observed no difference in the dynamics of the $\omega_0$ of the 2 eV oscillator (Fig. S10). Moreover, we restrict here our discussion to the effect of the pumps polarized along [100], but the experimental findings are the same for the other pump polarization (Fig. S9). We refer to Fig. S8 for a complete set of the measured reflectivity maps.

The first row in Fig. 6 reports – as a reference – the single-pump *SW* dynamics initiated by the (a) above- and (b) sub-gap excitation in the SC phase (reproduced from Fig. 2e,f). In the second row, we plot the *SW* dynamics obtained in the double-pump experiment: in c) we report the near-IR induced *SW* dynamics measured in the sample previously excited at t=-1 ps by the sub-gap pulse; in d) the order of the pumps is swapped, and we plot the mid-IR *SW* dynamics measured after the above-gap excitation.

The comparison of the *SW* dynamics photo-induced by the high-photon energy pulses with or without the previous interaction with the low-photon energy pump reveals that the dynamics is not significantly affected by the initial state of the sample, be it initially in equilibrium or previously driven into a non-thermal state by the excitation with the sub-gap pump (Fig. 6c). On the contrary, if the order of the two pumps is reversed (i.e. the near-IR pump prepares a non-thermal state that is then probed by the mid-IR pump-probe sequence) the *SW* dynamics is markedly different (Fig. 6d). When a QP population is excited by the above-gap pump, the $SW$ dynamics revealed by the sub-gap pump is qualitatively similar to the one measured by the near-IR pump only, featuring an initial fast-decaying response superimposed to a slower delayed recovery dynamics.

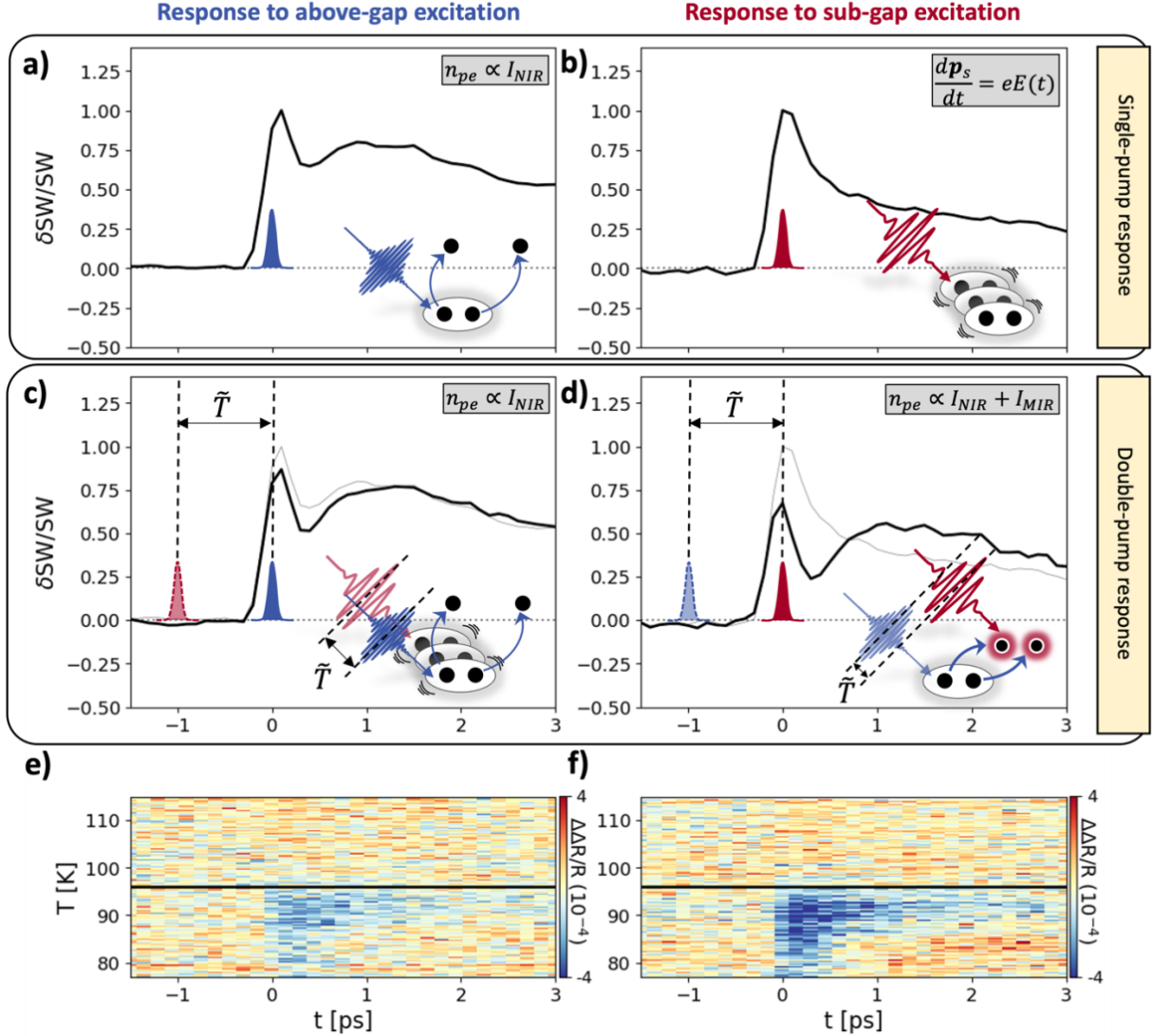

**Figure 6: Non-thermal quasiparticles in superconducting phase upon photoexcitation by above- and sub-gap pulses. a), b)** Normalized spectral weight dynamics extrapolated from the single-pump response to the above- and the sub-gap pulse, respectively (same as Fig. 2e,f). The polarization of the pumps is parallel to [100] and the sample temperature is 74 K. **c)** Spectral weight dynamics (black line) measured by the above-gap pulse when the system has been previously photoexcited by the mid-IR pulse at t=-1 ps (red Gaussian curve). **d)** Spectral weight dynamics (black line) measured by the sub-gap pulse in a sample that has previously (at t=-1 ps) interacted with the near-IR pulse. Grey curves in c) and d) are the single-pump spectral weight dynamics displayed in a) and b), over which the black curves have been renormalized. The drawings (bottom right in the panels) explain how the quasiparticle distribution is affected by the single- or the double-excitation. The equations (top rightpanels) summarize the main results of the kinetic model. **e), f)** Temperature-dependent differential signal obtained by subtracting the single-pump responses (in a) and b), respectively) from the double-pump response measured with both pumps (in c) and d), respectively). The time in the map (horizontal axis) is the delay between the second pump and the white-light probe, while $\tilde{T}$ (delay between the two pumps) is fixed to 1 ps.

Importantly, this difference is a peculiar property of the SC phase. In Fig. 6e,f, we plot the differential reflectivity ($\Delta\Delta R/R$) across the SC phase transition, obtained as the difference between the double-pump response in Fig 6c,d and the single-pump response in Fig. 6a,b for the near-IR and the mid-IR photoexcitation, respectively. In both cases, the differential signal vanishes above $T_C$ (dashed black line), indicating that the response to double-excitation is equivalent to the sum of the two single-pump experiments above the critical temperature.

## F. KINETIC-EQUATION DESCRIPTION OF THE ELECTRON DYNAMICS

**Theory**

In order to rationalize these results in a comprehensive framework, we resorted to the quasiclassical kinetic theory initially derived for conventional superconductors [36,37,66,67] and later extended to anisotropic gaps by Kabanov and co-workers [38]. The leading idea of the model is to describe the effect of the photoexcitation as a sudden injection of excess QPs that perturb the equilibrium distribution function. The subsequent QP recombination dynamics is governed by the emission and re-absorption of high-frequency phonons, while low-frequency phonons do not take part directly in the relaxation process. Importantly, the model relies on the assumption that the density of photoexcited QP is significantly smaller than the normal state carrier density, i.e. the photoexcitation is weak and does not fill the SC gap with states. In this respect, the pump fluence studies revealing a linear dependence of the transient signal on the pump intensity (Fig. S1, S2) ensure the validity of this assumption in our measurements. Formally, the kinetic equations describing the QP ($f_\epsilon$) and the phonon ($N_q$) distribution functions can be written in the form:

$$\begin{cases} \dfrac{\partial f_\epsilon}{\partial t} + \dfrac{\partial \epsilon_p}{\partial p}\dfrac{\partial f_\epsilon}{\partial r} - \dfrac{\partial \epsilon_p}{\partial r}\dfrac{\partial f_\epsilon}{\partial p} = I\{f_\epsilon\} + Q \\ \dfrac{\partial N_q}{\partial t} = I_{ph-ph}\{f_\epsilon\} + I_{ph-e}\{f_\epsilon\} \end{cases} \qquad \text{Eq. 1}$$

In the first equation, $\epsilon_p = \sqrt{\varepsilon_p^2 + \Delta_{SC}^2} + \boldsymbol{p}_s \boldsymbol{v}$ is the energy of the QP, where, in addition to the usual energy dispersion involving $\varepsilon_p = \dfrac{p^2}{2m} - \mu + \phi + \dfrac{p_s^2}{2m}$ and the SC gap $\Delta_{SC}$, we included the kinetic energy related to the movement of the condensate in its coordinate system [66]. In the notation used, $\mu$ is the chemical potential, $\phi$ is the electrostatic potential and $\boldsymbol{p}_s$ is the condensate momentum. We have denoted by $Q$ the source of QP in the system (i.e. the photoexcitation) and with $I\{f_\epsilon\} = I_{im} + I_{e-ph} + I_{ee}$ the collision integrals describing, respectively, collisions between QP and impurities, QP and phonons and electro-electron collisions. In the second equation, the first term on the right-hand side is the phonon-phonon collision integral for anharmonic processes and the second term the collision integral describing the collisions of phonons with Bogoliubov QP.

We will now see how, depending on the photon energy $h\nu$ of the excitation with respect to the amplitude of the antinodal SC gap, we can discriminate two cases, eventually leading to two different physical scenarios: $h\nu \gg 2\Delta_{SC}$, that is the case of the near-IR excitation in our experiment; $h\nu < 2\Delta_{SC}$, that corresponds to the mid-IR excitation.

The effect of high-energy fields ($h\nu \gg 2\Delta_{SC}$) has been widely studied in literature. Since the photon energy is larger than the antinodal gap, the high-energy pump is very efficient in creating QPs and subsequently breaking Cooper pairs; therefore, the leading term in Eq. 1 is $Q$. We sketched the expected pump-induced Cooper pair breaking in the bottom right of Fig. 6a. If the temporal duration of the pump is shorter than the relaxation processes, $Q$ determines the initial distribution of QPs. Furthermore, in the limit in which the electron-electron collisions and the electron-phonon recombination processes quickly lead to the phonon bottleneck [68], the distribution functions for phonons ($N_{\nu_q}$) and quasiparticles ($f_\epsilon$) can be respectively approximated as [38,69]:

$$N_{\nu_q} = \begin{cases} \dfrac{1}{e^{\frac{h\nu_q}{k_B T}} - 1}, & h\nu_q < 2\Delta_{SC} \\ \dfrac{1}{e^{\frac{h\nu_q}{k_B T^*}} - 1}, & h\nu_q > 2\Delta_{SC} \end{cases} \qquad \text{Eq. 2}$$

$$f_\epsilon = \frac{1}{e^{\frac{\epsilon}{k_B T^*}} + 1} \qquad \text{Eq. 3}$$

Where $T$ is the temperature of the low-frequency phonons ($h\nu_q < 2\Delta_{SC}$, i.e. the lattice temperature) and $T^*$ is the temperature of QP and high-frequency phonons ($h\nu_q > 2\Delta_{SC}$). This approximation, that is equivalent to the model introduced in Ref. [70], entails that the relaxation of the bottleneck is ruled by the escape of high-frequency phonons below the gap via phonon-phonon collisions, as in the Rothwarf-Taylor description [71]. By imposing the conservation of energy, the photoexcited QP density ($n_{pe}$) can be derived and, in the case of a temperature-dependent gap $\Delta(T)$ that best describes the recombination rates in optimally-doped cuprate samples [38], it is given by:

$$n_{pe} = \frac{\mathcal{E}_I / (\Delta(T) + k_B T/2)}{1 + \dfrac{2\nu}{N(0)\Omega_c}\sqrt{\dfrac{2k_B T}{\pi\Delta(T)}} e^{-\frac{\Delta(T)}{k_B T}}} \qquad \text{Eq. 4}$$

Where $\mathcal{E}_I$ denotes the absorbed energy density of the high-energy pump pulse, $\nu$ is the effective number of phonon modes per unit cell involved in the recombination processes, $N(0)$ is the density of states at the Fermi energy and $\Omega_c$ is the phonon frequency cut-off. The temperature dependence of the photoexcited QP density in Eq. 4 highlights two important aspects: i) the QP density $n_{pe}$ scales linearly with the intensity of the high-energy pump; ii) $n_{pe}$ is expected to approach zero when the temperature reaches T$_C$. The good agreement found in optimally-doped YBCO samples between theory and single-color 1.5 eV pump-probe transmission measurements confirms the efficacy of the model [38]. We included the expected linear dependence of the photoexcited QP density from the intensity of the near-infrared pump ($I_{NIR}$) in our experiment in the top right of Fig. 6a.

The case of low-energy fields ($h\nu < 2\Delta_{SC}$) is instead very different. Looking back at the first row in Eq. 1 and considering a spatially uniform photoexcitation (so that all gradients vanish, except for the gradient of the electrostatic potential), the effect of low-energy excitations enters the kinetic equation via the gradient in the third term on the left-hand side of Eq. 1. As sub-gap pulses do not deliver sufficient energy to break Cooper pairs, the number of QPs is conserved following the photoexcitation. As a consequence, the main effect of this pump is limited to the acceleration of the condensate, as depicted in the bottom right of Fig. 6b and formally described by the equation [66,69,72]:

$$\frac{d\boldsymbol{p}_s}{dt} = eE(t) \qquad \text{Eq. 5}$$

Where $E = -\nabla \phi$ is the electric field. In conventional *s*-wave superconductors, Eq. 5 leads to oscillation of the supercurrent, but it does not produce new QPs; at finite temperature, irradiation by sub-gap pulses can promote the already present thermally excited QPs to higher-energy empty states. This redistribution of the Bogoliubov excitations into a more favourable non-equilibrium distribution can eventually lead to an enhancement of superconductivity, known as the "Eliashberg effect" [73], that has been experimentally observed in superconducting thin films irradiated by THz pulses [74].

It should be highlighted, however, that this description might differ in the case of *d*-wave superconductors: pumps with photon energy smaller than the antinodal gap can still create QPs in the nodal directions. However, this effect is rather small, as the QP density of states within the gap is substantially suppressed, leading to a smaller probability of Cooper pair-breaking processes.

**Simulations**

The effect of the photoexcitation with sub-gap pulses does therefore depend on the initial state of the sample. If the sample is in equilibrium before the excitation, the dynamics is ruled by the thermal distribution of QPs. In Fig. 7a, we simulate the momentum- and temperature-resolved QP population in equilibrium, obtained by multiplying the *d*-wave QP density of states [75] by the Fermi-Dirac distribution. In the calculations, we have parametrized the gap energy dispersion in momentum space as $\Delta(k) = \Delta_{SC} \sin(ka)$, where $\Delta_{SC}$ is the amplitude of the antinodal gap. In this parametrization, $ka = 0$ indicates the nodal direction and $ka = \pi/2$ the antinodal direction. As expected, at low temperatures $T/T_C \ll 1$, thermal QPs are located only at low energies, namely at the nodes where $ka \sim 0$. As soon as the temperature increases and approaches $T_C$, thermal QPs populate states at higher energies and move away from the node. However, even at $T_C$, thermal excitations are not sufficient to populate the antinodal states. In the simulations we have considered $\Delta_{SC}$=75 meV and $T_C$ = 90 K.

In the three-pulse experiment, however, the mid-IR pump impinging on the sample does not interact with an equilibrium QP distribution, but instead with a non-thermal QP population previously injected by the near-IR pump according to Eq. 4. In order to introduce the momentum-dependence of the gap amplitude, we rescaled the temperature-dependent gap in Eq. 4 as a function of $k$, so that $\Delta(k, T) = \Delta_{SC} \sin(ka) \tanh\left(1.75 \times \sqrt{\frac{T_C}{T} - 1}\right)$. We highlight that this procedure, even though neglects the *k*-dependent phonon distribution, is in general reliable since in the scattering events between QPs and impurities momentum is conserved. The resulting photoexcited QP density $n_{pe}$ is plotted in Fig. 7b as a function of temperature and momentum. The simulation shows that photo-injected QPs populate states at all momenta, in agreement with previous ARPES findings [26,28].

The comparison between the equilibrium (Fig. 7a) and the non-thermal QP distribution (Fig. 7b) reveals that antinodal states can be populated only upon photoexcitation by an above-gap pump. Furthermore, as displayed in the temperature cuts in Fig. 7b, the creation of antinodal QPs is expected at all temperature below $T_C$.

The two distinct scenarios in Fig. 7a,b can be singled out in the three-pulse experiment, where the order of arrival of the pumps discriminates two cases.
1) If the mid-IR pump impinges before the near-IR one (Fig. 6c), the first pulse accelerates the condensate according to Eq. 5 and creates few low-energy ($\epsilon \ll \Delta_{SC}$) excitations. Since the sample is at non-zero temperature, the response is governed by absorption, i.e., by thermally-excited QPs (Fig. 7a). When the second near-IR pulse arrives 1 ps later, additional QPs are created at higher energy and momentum (Fig. 7b). The total response will therefore be that of the second pump only and can be described by Eq. 4, in which the photoexcitation intensity $\mathcal{E}_I$ is that of the near-IR pump alone, $I_{NIR}$. This is summarized in the sketches in the bottom of Fig. 6c.
2) When the order of arrival is swapped (Fig. 6d), however, there is a substantial amount of antinodal QPs photoexcited by the first impinging near-IR pump, as shown in Fig. 7b. This non-equilibrium QP distribution can absorb energy from the mid-IR pump, that – when impinging on the sample 1 ps later - can now couple to the previously excited non-thermal QPs. Again, the photoresponse is determined by Eq. 4, where $\mathcal{E}_I$ is now the sum of the absorbed energies from the first and the second pump, $I_{NIR} + I_{MIR}$. The dynamics measured is thus qualitatively similar to the one initiated by the photoexcitation with high photon energy alone, as observed in the extrapolated *SW* dynamics.

The fact that the differential response vanishes above the critical temperature (Fig. 6e,f) is a further confirmation that the difference found between high- and low-photon energy excitations is closely related to the presence of the SC gap. The closure of the gap at $T_C$ suppresses the *d*-wave anisotropy and, in turn, the difference in the *k*-dependent QP distribution excited by the two pumps.

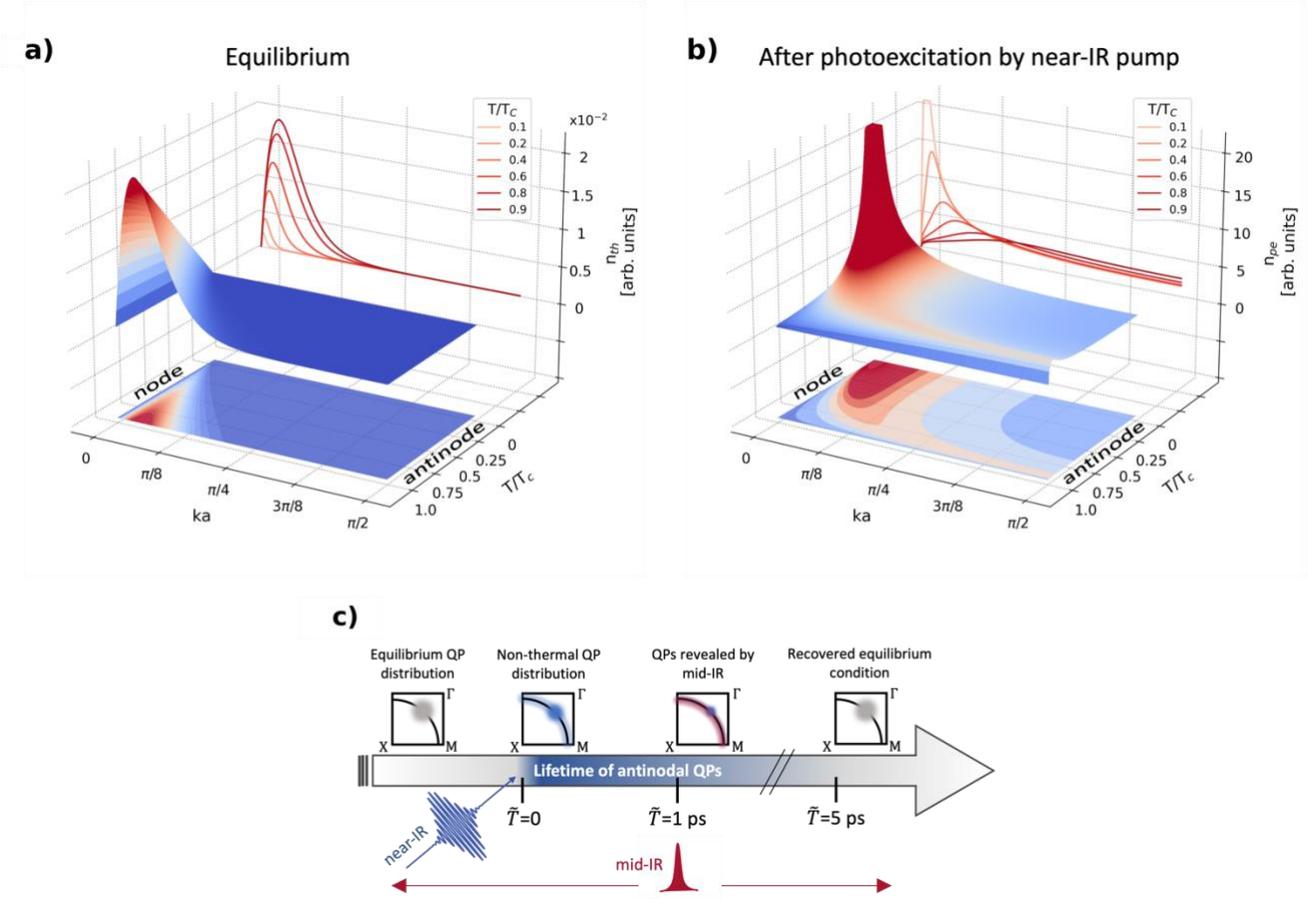

**Figure 7: Simulated momentum- and temperature-resolved quasiparticle distributions. a)** Equilibrium quasiparticle density ($n_{th}$) obtained by multiplying the QP density of states of a *d*-wave superconductor [75] by the quasiparticle Fermi-Dirac distribution. The red curves are cuts as a function of momentum at different temperatures listed in the legend. **b)** Photoexcited quasiparticle density ($n_{pe}$) according to Eq. 4 for a temperature- and momentum-dependent gap $\Delta(k,T)$. For visual needs, the distribution has been cut at $n_{pe}$=20. As showed in the red cuts, an amount of antinodal quasiparticles is photoexcited at all temperatures. **c)** Sketch of the lifetime of the photoexcited antinodal QPs revealed by the mid-IR three-pulse spectroscopy.

To conclude, in the three-pulse experiment the interaction with the mid-IR pump represents a means to probe the presence of antinodal QPs. Indeed, the response to sub-gap pulses is different whether or not the antinodal regions have been populated with QPs by the previous interaction with a high-energy pump, thus leading to a non-zero differential signal in the SC phase (Fig. 6f).

The mid-IR pump can then be used to probe the lifetime of the antinodal QP population. By tuning the delay $\tilde{T}$ between the two pumps, it is possible to track the recombination dynamics of previously excited QPs, that is pictorially illustrated in Fig. 7c. At $\tilde{T} < 0$, mid-IR pulses interact with a thermal population of QPs located at the nodal region of the first Brillouin zone (leftmost sketch in Fig. 7c). When the photoexcitation by the near-IR pump suddenly injects antinodal QPs, a non-null differential response emerges at $\tilde{T} > 0$. Importantly, by quantifying the differential response as a function of $\tilde{T}$, the lifetime of antinodal QPs can be extrapolated. In Fig. S11, we plot the integrated differential response below $T_C$, showing that it displays an exponentially decaying dynamics and vanishes at $\tilde{T} \sim 5$ ps. By fitting the data, we estimate a decay time of ~ 1.2 ps. This observation matches well the reported dynamics of the photo-induced gap filling in Bi2212 that has been related to the ultrafast melting of the condensate driven by phase fluctuations [57]. This result, which hints at a connection between the recombination rate of QPs and the recovery dynamics of phase coherence, corroborates previous findings [76] and calls for future experimental and theoretical investigations.

## IV. CONCLUSIONS

We have investigated the optical response of optimally-doped $Bi_2Sr_2Ca_{0.92}Y_{0.08}Cu_2O_{8+\delta}$ to photoexcitation both above and below the antinodal gap. By modelling the broadband transient reflectivity with a time-dependent Drude-Lorentz dielectric function, we found that the dynamics is ruled by the transient modification of an effective excitation at 2 eV. Our measurements do not clarify the nature of this transition, which, based on previous reports, could involve both in-plane Cu-O charge-transfer excitations [2,46] and *dd* orbital transitions [47,50]. In this respect, we highlight that the second scenario, which is usually deemed unlikely because of the optical inactivity of the *dd* transitions, might be enabled by phonon-assisted processes which lift the optical selection rules, as observed in other copper oxides [77].

Interestingly, while the pump-induced shift of this interband excitation is independent of all experimental parameters (pump photon energy, pump polarization and temperature), the dynamics of its spectral weight strongly depends on the properties of the pump pulse. Moreover, the sign change of the transient spectral weight at the critical temperature indicates a connection between this observable and the presence of the superconducting order. In contrast, the absence of a clear discontinuity at $T_C$ in the dynamics of the central frequency might suggest that the transient shift of the high-energy excitation is related instead to the formation of other local electronic gaps. Further studies are required to test the universality of this observation.

The simultaneous excitation with both near-IR and mid-IR pumps combined in a three-pulse scheme allowed us to show that the quasiparticle dynamics triggered by above- and below-gap photoexcitations are different in the superconducting phase, as a result of the anisotropy of the gap. In the presence of a *d*-wave gap, while near-IR pulses deliver sufficient energy to target the entire Fermi surface and accumulate non-thermal quasiparticles in the antinodal states that are inaccessible in equilibrium, mid-IR pulses mainly affect the nodal regions and accelerate the condensate. Above the critical temperature, this dichotomy vanishes, and the momentum-selectivity of the mid-IR excitation is suppressed. The mid-IR based three-pulse technique allows us to capture the recombination dynamics of photo-injected quasiparticles, that is compatible with previous time-resolved ARPES studies. Our findings are consistent with the kinetic theory for superconductors [38] and highlight that excitation by pulses with suitable photon energy may provide a means to understand and control the non-equilibrium distribution of quasiparticles in momentum space.

**ACKNOWLEDGEMENTS**

The authors acknowledge fruitful discussions with Stefano Dal Conte and Jeffrey Davis. This work was supported by the European Commission through the projects INCEPT (ERC-2015-STG, Grant No. 677488) and COBRAS (ERC-2019-PoC, Grant No. 860365). D.F. acknowledges funding from the MIUR through the PRIN program No. 2017BZPKSZ. The work at the University of Minnesota was funded by the U.S. Department of Energy through the University of Minnesota Center for Quantum Materials, under Grant No. DE-SC0016371. This research was undertaken thanks in part to funding from the Max Planck-UBC-UTokyo Centre for Quantum Materials and the Canada First Research Excellence Fund, Quantum Materials and Future Technologies Program. This project is also funded by the Natural Sciences and Engineering Research Council of Canada (NSERC); the Alexander von Humboldt Fellowship (A.D.); the Canada Research Chairs Program (A.D.); and the CIFAR Quantum Materials Program.

# Dynamics of non-thermal states in optimally doped Bi$_2$Sr$_2$Ca$_{0.92}$Y$_{0.08}$Cu$_2$O$_{8+\delta}$ revealed by mid-infrared three-pulse spectroscopy

Angela Montanaro[1,2,3], Enrico Maria Rigoni[1,2], Francesca Giusti[1,2], Luisa Barba[4], Giuseppe Chita[4], Filippo Glerean[5], Giacomo Jarc[1,2], Shahla Y. Mathengattil[1,2], Fabio Boschini[6,7], Hiroshi Eisaki[8], Martin Greven[9], Andrea Damascelli[7,10], Claudio Giannetti[11,12], Dragan Mihailovic[13], Viktor Kabanov[13], and Daniele Fausti[1,2,3*]

[1]*Department of Physics, Università degli Studi di Trieste, 34127 Trieste, Italy*
[2]*Elettra Sincrotrone Trieste S.C.p.A., 34149 Basovizza Trieste, Italy*
[3]*Department of Physics, University of Erlangen-Nürnberg, 91058 Erlangen, Germany*
[4]*Institute of Crystallography, CNR, Elettra Sincrotrone Trieste S.C.p.A., 34149 Basovizza Trieste, Italy*
[5]*Department of Physics, Harvard University, Cambridge, Massachusetts 02138, USA*
[6]*Centre Énergie Matériaux Télécommunications, Institut National de la Recherche Scientifique, Varennes, Québec, Canada J3X1S2*
[7]*Quantum Matter Institute, University of British Columbia, Vancouver, BC, Canada V6T 1Z4*
[8]*Nanoelectronics Research Institute, National Institute of Advanced Industrial Science and Technology, Tsukuba, Ibaraki 305-8568, Japan*
[9]*School of Physics and Astronomy, University of Minnesota, Minneapolis, Minnesota 55455, USA*
[10]*Department of Physics & Astronomy, University of British Columbia, Vancouver, BC, Canada V6T 1Z1*
[11]*Department of Mathematics and Physics, Università Cattolica, I-25121 Brescia, Italy*
[12]*Interdisciplinary Laboratories for Advanced Materials Physics (I-LAMP), Università Cattolica, I-25121 Brescia, Italy*
[13]*Jožef Stefan Institute, Jamova 39, 1000 Ljubljana, Slovenia*

*Correspondence: daniele.fausti@elettra.eu*

# SUPPLEMENTARY MATERIALS

**NOTE S1.** Optical properties at equilibrium
**NOTE S2.** Differential fit in the superconducting phase
**NOTE S3.** Differential fits in the normal and the pseudogap phases

**FIGURE S1.** Fluence dependence of the mid-infrared pump
**FIGURE S2.** Fluence dependence of the near-infrared pump
**FIGURE S3.** Fluence dependence of the probe
**FIGURE S4.** Differential fits at high fluence in the superconducting phase
**FIGURE S5, S6.** Temperature-dependent reflectivity maps
**FIGURE S7.** Differential fit of the temperature maps
**FIGURE S8.** Double-pump measurements in the superconducting phase
**FIGURE S9.** Spectral weight dynamics in the double-pump experiment ([110] axis)
**FIGURE S10.** Central frequency dynamics in the double-pump experiment
**FIGURE S11.** Double-pump dynamics

# NOTE S1. Optical properties at equilibrium

The starting point of the differential fits used to model the transient response in our pump-probe measurements is the reflectivity of the sample at equilibrium. We have used the ellipsometry data published in [1,2], in which the authors measured the dielectric function and the static reflectivity of an optimally doped Y-Bi2212 sample at three different temperatures (20, 100 and 300 K), corresponding to the superconducting, pseudogap and normal phases, respectively. We have considered a Drude-Lorentz model to fit the dielectric function. Unlike Refs. [1,3], we have used a simple Drude peak instead of an extended Drude model to fit the low-energy side of the spectrum, as the main focus of our study is the optical response in the visible region overlapping with our broadband probe and not the low-energy mid-IR transitions. We used the Reffit tool [4] to obtain a Kramers-Kronig constrained fit to the data, using the following equations:

$$\varepsilon_1(\omega) = \varepsilon_\infty + \sum_{i=1}^{N} {\omega_P^i}^2 \left[ \frac{{\omega_0^i}^2 - \omega^2}{\left({\omega_0^i}^2 - \omega^2\right)^2 + {\gamma^i}^2 \omega^2} \right]$$

$$\varepsilon_2(\omega) = \sum_{i=1}^{N} {\omega_P^i}^2 \left[ \frac{\gamma^i \omega}{\left({\omega_0^i}^2 - \omega^2\right)^2 + {\gamma^i}^2 \omega^2} \right]$$

where $\varepsilon_1(\omega)$ and $\varepsilon_2(\omega)$ are the real and imaginary part, respectively, of the energy-dependent dielectric function. The sum over the $i$-index runs over the minimum number of oscillators ($N$) needed to fit the data. For each oscillator, the central frequency ($\omega_0$), the plasma frequency ($\omega_P$) and the scattering rate ($\gamma$) are estimated. The Drude peak is included in the sum as an oscillator at $\omega_0=0$ and this parameter is kept fixed during the fitting procedure. The phenomenological parameter $\varepsilon_\infty$ accounts for the higher-energy bands outside the measured range and, for consistency, has also been kept fixed and equal to the one given by [1].

We plot in Fig. A the energy-dependent reflectivity (logarithmic scale) at the three temperatures obtained through the Drude-Lorentz fitting and calculated as:

$$R(\omega) = \left| \frac{1 - \sqrt{\varepsilon(\omega)}}{1 + \sqrt{\varepsilon(\omega)}} \right|^2$$

where $\varepsilon(\omega)$ is the complex dielectric function. The three curves in Fig. A overlap on the high-energy scale and differ mainly below the charge-transfer (CT) edge (<2 eV).

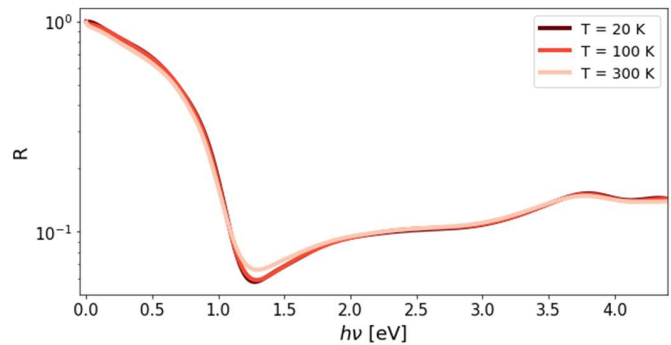

**Figure A: Static reflectivity at different temperatures (logarithmic scale).** Raw data taken from Refs. [1,2] and fitted via a Drude-Lorentz model. The best fitting parameters are reported in Table A.

We summarize in Table A the best fitting parameters in the three phases. In order to fit the low-energy side of the spectrum within a simple (not-extended) Drude model, we have included two mid-IR peaks, whose parameters have not been constrained. These oscillators account for the intraband transitions and are the most affected (along with the Drude peak) by the sample temperature. The visible region of the reflectivity is modelled by three oscillators whose central frequencies are $\omega_0 \sim 1.46$, 2 and 2.72 eV in the superconducting phase. Three higher-energy contributions (>3 eV) model the dielectric function above the CT-edge.

**Table A:** Drude-Lorentz parameters obtained via Kramers-Kronig constrained fits [4] of the static reflectivity at different temperatures (superconducting, pseudogap and normal phase). All values are expressed in [cm$^{-1}$]. The red values have been kept fixed during the fitting procedure.

| | | T=20K | T=100K | T=300K |
|---|---|---|---|---|
| Drude peak | $\epsilon_\infty$ | 2.67 | 2.62 | 2.62 |
| | $\omega_0$ | 0 | 0 | 0 |
| | $\omega_P$ | 10766 | 11492 | 10628 |
| | $\gamma$ | 26 | 180 | 349 |
| Mid-infrared peaks | $\omega_0$ | 942 | 1140 | 1345 |
| | $\omega_P$ | 13411 | 13928 | 14654 |
| | $\gamma$ | 3985 | 5532 | 5526 |
| | $\omega_0$ | 6309 | 6530 | 6626 |
| | $\omega_P$ | 5102 | 2279 | 1174 |
| | $\gamma$ | 4841 | 2958 | 1707 |
| Visible peaks | $\omega_0$ | 11800 | 12305 | 12305 |
| | $\omega_P$ | 2358 | 2261 | 2472 |
| | $\gamma$ | 3644 | 4872 | 4872 |
| | $\omega_0$ | 16163 | 16075 | 16026 |
| | $\omega_P$ | 6385 | 5296 | 5500 |
| | $\gamma$ | 8304 | 8160 | 8097 |
| | $\omega_0$ | 21947 | 21947 | 21947 |
| | $\omega_P$ | 15026 | 15600 | 15603 |
| | $\gamma$ | 13998 | 13998 | 14688 |
| Higher-energy contributions | $\omega_0$ | 31165 | 31059 | 30913 |
| | $\omega_P$ | 17039 | 17363 | 18115 |
| | $\gamma$ | 5896 | 6187 | 6884 |
| | $\omega_0$ | 35567 | 35479 | 34598 |
| | $\omega_P$ | 17317 | 16448 | 11610 |
| | $\gamma$ | 5765 | 6224 | 5758 |
| | $\omega_0$ | 41677 | 40614 | 39715 |
| | $\omega_P$ | 27979 | 27430 | 28795 |
| | $\gamma$ | 5000 | 5000 | 6949 |

# NOTE S2. Differential fit in the superconducting phase

In this section, we describe in detail how the differential fits of the transient broadband reflectivity maps have been performed. As a reference, we will only consider the reflectivity map in Fig. 2d in the main ($h\nu_{pump}$=1.44 eV, pump polarization parallel to the [100] axis). The procedure is the following: we compute the static reflectivity ($R_{eq}(\omega)$) at a given temperature obtained through the best-fitting Drude-Lorentz model in Table S1. We duplicate the reflectivity and create a new function ($R_{exc}(\omega,t)$) in which we let selected parameters to change at different pump-probe delays. We finally directly fit the measured $\Delta R/R(\omega,t)$ map using the following equation: $(R_{exc}(\omega,t) - R_{eq}(\omega))/R_{eq}(\omega)$. In this way, the fit returns the pump-induced change in the oscillator parameters that have been modified to fit the transient data.

In Fig. B,a we illustrate the fit results for a selected time delay (t=200 fs) for the transient reflectivity upon photoexcitation by the near-infrared pump polarized along [100]. The black points indicate the data with the relative error bars, while the red line is the differential fit obtained. We report the $\chi^2_{red}$ associated to the corresponding fit on the upper left corner of the panel and the best fitting parameters (and their uncertainties) on the lower right. In each panel, we modify a different oscillator in the Drude-Lorentz model, as reported in the title. In contrast to the signal in the pseudogap and the normal phase, an impulsive modification of the scattering rate of the Drude peak cannot describe the data in the superconducting phase (top left panel). In the following panels, we have modified at the same time the three parameters ($\omega_0, \omega_P, \gamma$) that fully characterize each oscillator. The modification of the mid-IR peaks and the 1.46 eV one returns fairly good agreement on the low-energy side of our probing window, but it does not provide a good description at higher energies. The two best fits are obtained by modifying either the 2 eV oscillator or the 2.72 eV one.

In order to choose the most reliable description, we have calculated how the reflectivity would dynamically change in these two scenarios on the entire energy axis of the static measurements. In fact, being our detection bandwidth about 0.6 eV wide, a good fit in our probing window could result in strong (larger than the measured $\Delta R/R$) reflectivity variations even outside this region. We have therefore fitted the whole transient reflectivity map (i.e. for all the time delays, and not just a single temporal cut as in Fig. B,a) and obtained a "perturbed" model at each time step ($M_{exc}(t)$). We then used $M_{exc}(t)$ to compute the time-dependent reflectivity over the entire energy range. In Fig. B,b we show the results of this analysis. The panel on the left (right) is obtained by modifying the parameters of just the 2 eV (2.72 eV) oscillator. The solid black curve is the equilibrium reflectivity, while the dashed yellow curve is the computed out-of-equilibrium reflectivity at the pump-probe overlap (t=0). The color-coded lines indicate the relative reflectivity variation at different pump-probe delays (blue/negative times, red/positive times) multiplied by a $10^3$ factor. It is clear that, while modifying the oscillator centered at 2 eV results in a $10^{-3}$ reflectivity change that is mostly localized across our probing window, the time-dependent modification of the 2.72 eV oscillator mostly affects the higher-energy scale and results in a reflectivity change which is 10 times larger than the one we measured. For this reason, we will restrict our analysis to a pump-induced variation of the 2 eV oscillator only, which guarantees, at the same time, a good differential fit to our data and a more reasonable behaviour outside the detection window.

In Fig. C,a (left panel) we plot the degrees of correlation among the fitted ($\omega_0, \omega_P, \gamma$) parameters of the 2 eV oscillator. At each time-delay, we have performed a differential fit as described above. Each point in the plot is, at a given time-delay, one of the off-diagonal elements of the best-fitting symmetric 3x3 covariance matrix. The light green, yellow, dark green lines indicate the degree of correlation between $\{\omega_0,\omega_P\}, \{\omega_0,\gamma\}, \{\omega_P,\gamma\}$, respectively. It is clear that the three parameters in the fit are not independent and, in particular, $\omega_P$ and $\gamma$ display the highest degree of correlation. For this reason, we have tried to fit the data by modifying only two of the three parameters of the 2 eV oscillator. In the right panel of Fig. C,a we plot in light green (yellow, dark green) the degree of correlation computed by keeping $\omega_0$ ($\omega_P, \gamma$) fixed in the differential fit (i.e. the only off-diagonal element of the symmetric 2x2 covariance matrix). A joint analysis of the $\chi^2_{red}$ and the degree of parameters' correlation obtained in the three models led us to keep $\omega_P$ a fixed parameter in the fit. This choice, while does not significantly alter the quality of the fit, guarantees a degree of correlation which is two or even three orders of magnitude lower than in the other two scenarios (Fig. C,a right panel). We show in Fig. C,b-c the comparison between the data (as in Fig. 2 in the main) and the corresponding differential fits obtained by keeping $\omega_P$ fixed.

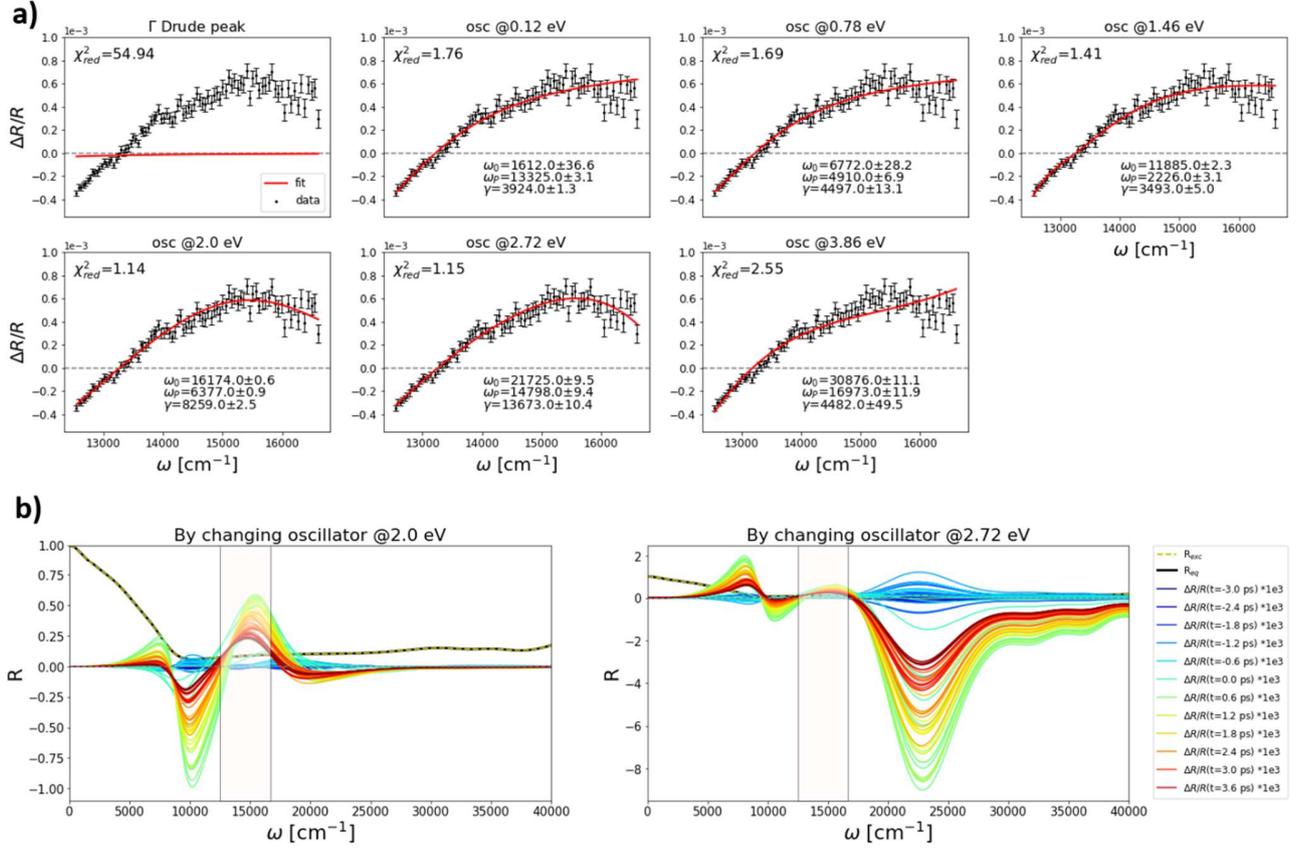

**Figure B: Differential fits in superconducting phase. a)** The black dots are energy-dependent reflectivity cuts at a fixed time-delay (t=200 fs) upon photoexcitation by the 1.44 eV pump polarized along [100] at T=74 K. The red line in each panel is a Drude-Lorentz differential fit obtained by modifying one oscillator at a time (reported in the title of each panel). The reduced $\chi^2_{red}$ and the best-fitting parameters are reported in the upper left and lower right corner in each panel, respectively. **b)** Static reflectivity (black curve), out-of-equilibrium reflectivity at the pump-probe overlap (yellow-dashed curve) and relative time-dependent reflectivity changes calculated over a broad frequency axis obtained by modifying the 2 eV (left panel) or the 2.72 eV (right panel) oscillator. The white-shaded area indicates are probing energy window.

The conclusion that we draw is that a Drude-Lorentz differential analysis is not able discriminate between an impulsive change in the plasma frequency or in the scattering rate of the oscillator. This is due to the fact that we do not impose any normalization constraint on the time-dependent fitting Lorentzian functions. It is important to highlight, however, that the $\omega_0$ dynamics is not affected by this choice and the central frequency of the oscillator remains an independent parameter of the differential fit.

A more convenient choice to analyse the time-dependent results of the differential fit is therefore to find an observable which is sensitive to the modification of either the plasma frequency or the scattering rate of the oscillator. We have used the spectral weight (*SW*) defined as follows:

$$SW = \int_0^\infty d\omega\, \sigma_1(\omega) = \frac{1}{4\pi}\int_0^\infty d\omega\, \omega\, \varepsilon_2(\omega)$$

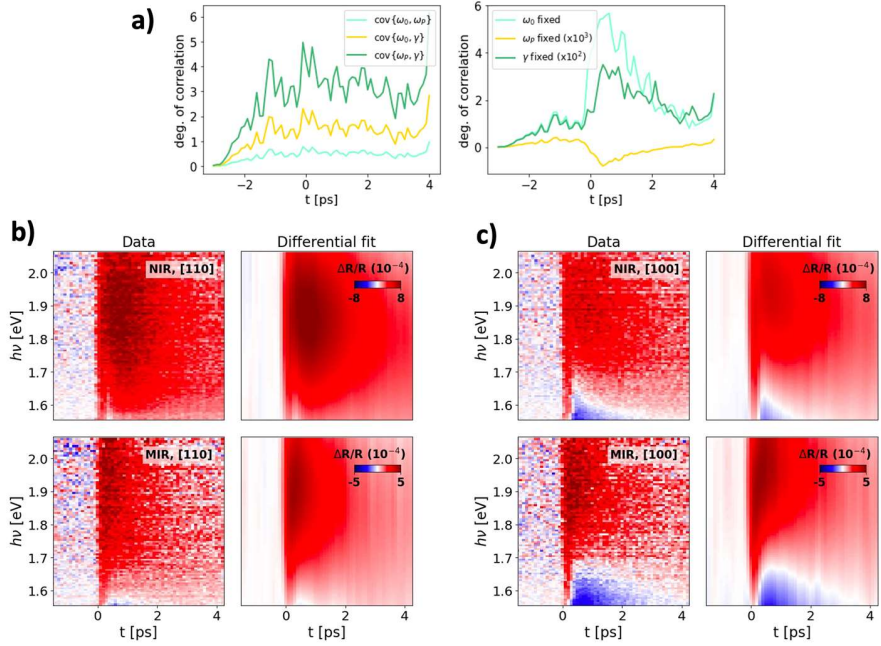

**Figure C: Differential fits with constrained parameters. a)** Left panel: off-diagonal elements as function of time-delay of the 3x3 covariance matrix obtained by modifying all the three parameters of the 2 eV oscillator. Each color refers to a different pair of parameters, as indicated in the legend. Right panel: time-dependent off-diagonal element of the 2x2 covariance matrix obtained by keeping fixed one of three parameters of the 2 eV oscillator. **b-c)** Comparison between data and fit keeping $\omega_P$ a fixed parameter. The labels in right upper corner of each reflectivity map indicate the photon energy and the polarization of the pump employed. The diverging colormap has been chosen to highlight where sign changes occur (white color).

where $\sigma_1(\omega)$ is the real part of the optical conductivity. In this framework, a pump-induced change in spectral weight ($\delta SW(t)$) is computed as the difference between the spectral weight calculated using the best-fitting "perturbed" model $M_{exc}(t)$ ($SW_{exc}(t)$)) and the one calculated using the equilibrium model $M_{eq}$ ($SW_{eq}$). The integration is computed over the whole energy axis of the static ellipsometry data in Fig. A.

We will discuss the results of the fits in terms of $\delta\omega_0/\omega_0(t)$ and $\delta SW/SW(t)$, that is the relative time-dependent change in the central frequency and in the spectral weight, respectively.

# NOTE S3. Differential fits in the normal and the pseudogap phases

Contrarily to the signal in the superconducting phase, the transient broadband reflectivity measured in either the normal or the pseudogap one can be fitted by modifying only the Drude peak in the dielectric function, without invoking any impulsive modification of the interband oscillators. This result further corroborates the fact that the spectral weight on the high-energy scale undergoes a redistribution only when the sample enters the superconducting phase.

We plot in Figure D the comparison between the reflectivity maps and the data, at T=300 K (from a) to d)) and at T=105 K (from e) to h)). As a reference, we consider here only the reflectivity map upon photoexcitation by the mid-IR pump polarized along [110]. We plot the result of the fit in Fig. D,d-h for the normal and pseudogap phases, respectively. At T=300 K, the pump induces a transient increase of the Drude scattering rate ($\Gamma$), that then reaches a positive plateau within a few ps. In the pseudogap phase, the photoexcitation has an opposite effect, resulting in a transient decrease of the scattering rate, and thus an increase of conductivity, that has already been observed in literature and has been ascribed to a partial photo-induced delocalization of the Mott-like localized states at equilibrium [5]. A subsequent scattering rate increase is observed after approximately 500 fs, that is compatible with the thermal heating of the sample.

It is worth to note that the dynamics discussed above are independent from the pump photon energy and the pump polarization. In Fig. D,d-h the four curves corresponding to different pump wavelength and polarization nearly overlap, apart from a scaling factor that could be due to a slightly difference in the pumping fluences used. This, together with the double-pump measurements (Fig. 5 in the main text), further confirms that the *k*-selectivity of the mid-IR pump photoexcitation is lost above the critical temperature, as a consequence of the closure of the superconducting gap

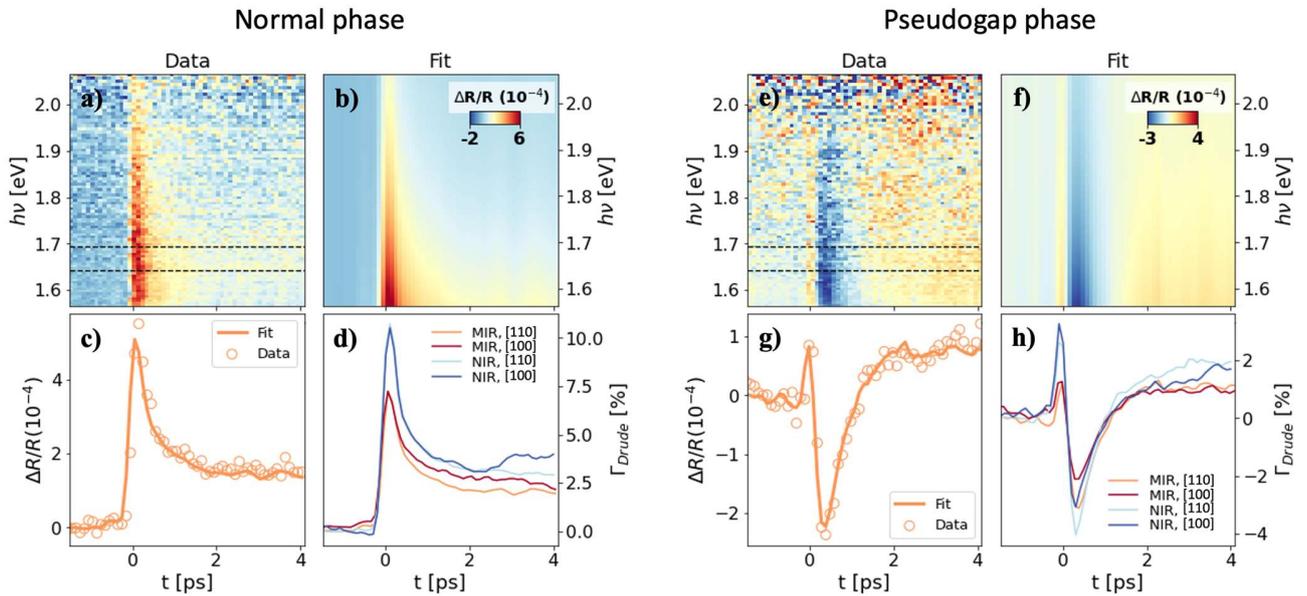

**Figure D: Differential fits in the normal and pseudogap phases. a-b)** Reflectivity map and corresponding differential fit in the normal phase (T=300 K) upon photoexcitation by a mid-IR pump pulse polarized along [110]. **c)** The open circles are a temporal cut of the map in a) averaged over the region within the dashed black lines. The solid line is the corresponding differential fit. **d)** Relative transient change of the Drude scattering rate ($\Gamma$) in the four configurations of pump photon energy (mid-IR or near-IR) and pump polarization (polarized along [110] or [100]). **e-h)** Same as a-d) but in the pseudogap phase (T=105 K).

# FIGURE S1. Fluence dependence of the mid-infrared pump

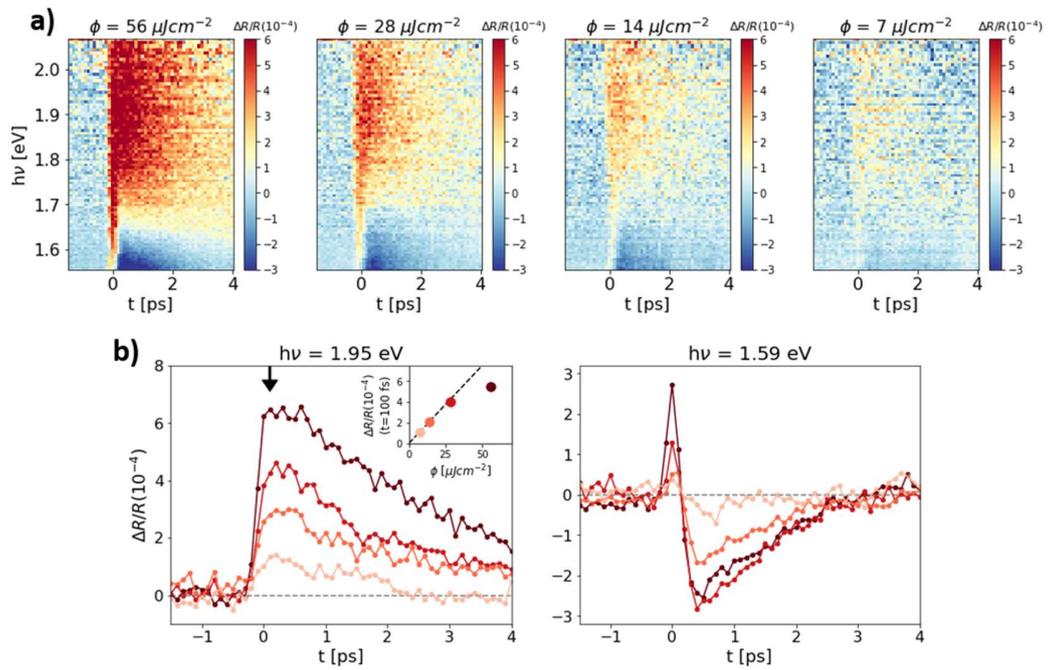

**Figure S1: a)** Color-coded transient reflectivity maps as function of probe energy and pump-probe delay for different mid-IR pumping fluences (pump polarized along [100]). **b)** Horizontal cuts (averaged over 50 meV) of the maps in a) at hν = 1.95 eV for all the fluences. Inset: fluence dependence of the signal at t=100 fs, as indicated by the black arrow. **c)** Same as b), but for hν = 1.59 eV.

# FIGURE S2. Fluence dependence of the near-infrared pump

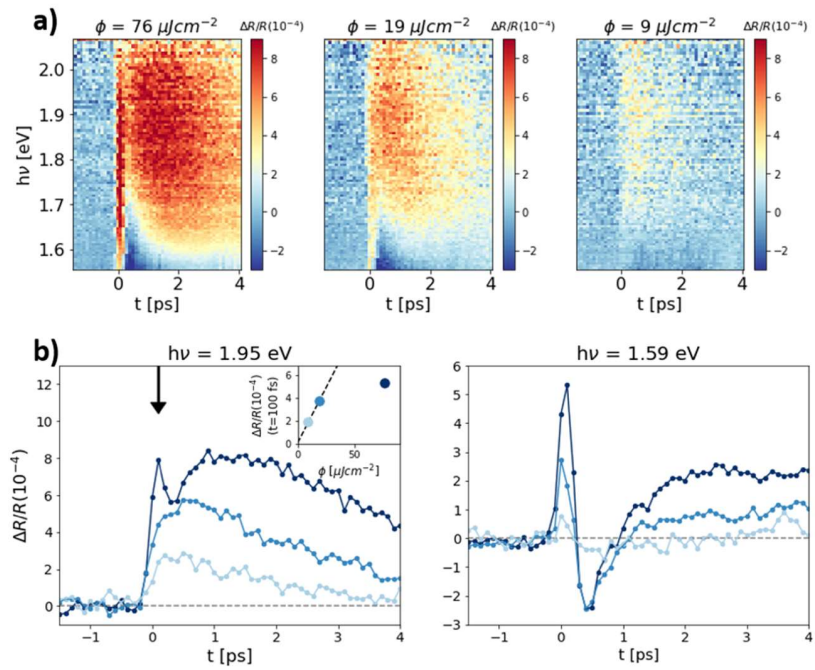

**Figure S2: a)** Time- and energy-dependent transient reflectivity upon photoexcitation by near-infrared pulses (polarized along [100]) at different fluences. **b)** Horizontal cuts of the maps in a) at hν=1.95 eV. Inset: fluence dependence of the signal at t=100 fs. **c)** Same as b), but for hν=1.59 eV.

# FIGURE S3. Fluence dependence of the probe

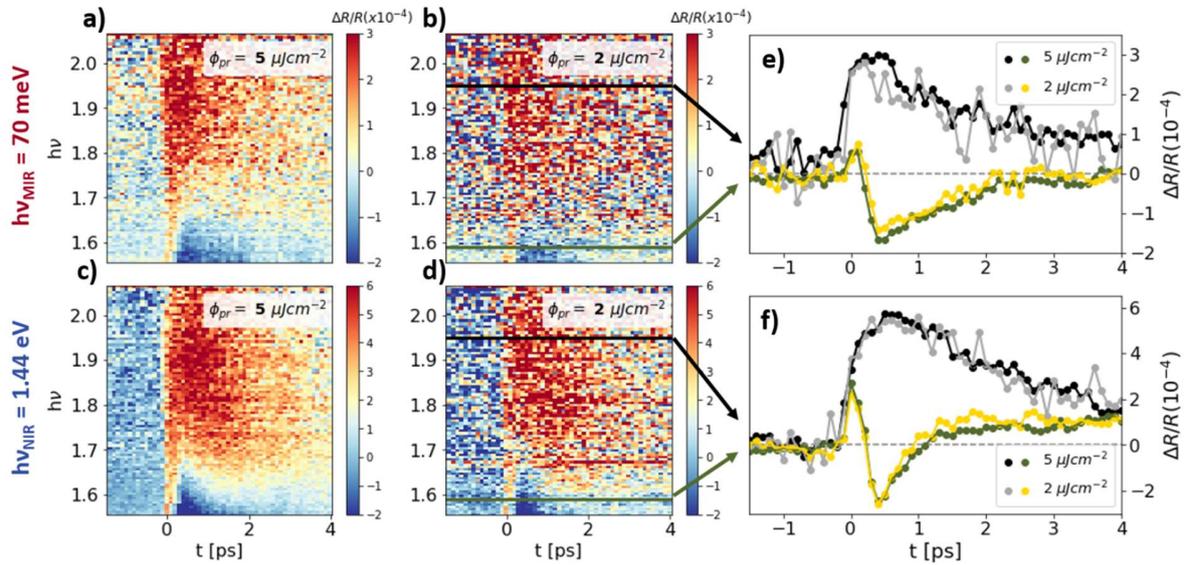

**Figure S3: a-b)** Reflectivity maps upon photoexcitation by a mid-IR pump pulse ($\phi_{MIR}$=14 µJ/cm$^2$) polarized along [100] for a probe fluence equal to 5 and 2 µJ/cm$^2$, respectively. **c-d)** Same as a-b) but for a near-IR pump pulse ($\phi_{NIR}$=19 µJ/cm$^2$). **e)** The black (green) and the grey (yellow) lines are the dynamics at hν = 1.95 eV (hν = 1.59 eV) of the mid-IR induced reflectivity maps at 5 and 2 µJ/cm$^2$, respectively. **f)** Same as e) but for the near-IR induced reflectivity maps.

# FIGURE S4. Differential fits at high fluence in the superconducting phase

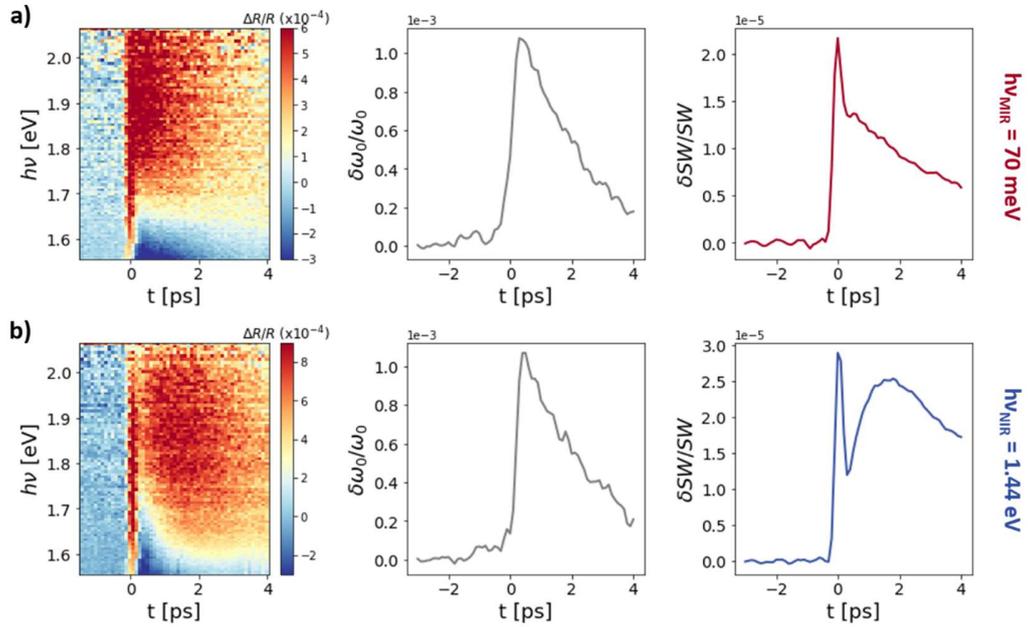

**Figure S4: a)** Left panel: time- and spectrally-resolved reflectivity induced by a mid-IR pump pulse polarized along the [100] ($\phi_{MIR}$ = 56 µJ/cm$^2$). Middle and right panels: results of the differential fit as function of the pump-probe delay in terms of the relative shift of the central frequency and the relative change in the spectral weight, respectively. **b)** Same as a) but for a near-infrared pump pulse ($\phi_{NIR}$ = 76 µJ/cm$^2$).

.

# FIGURE S5, S6. Temperature-dependent reflectivity maps

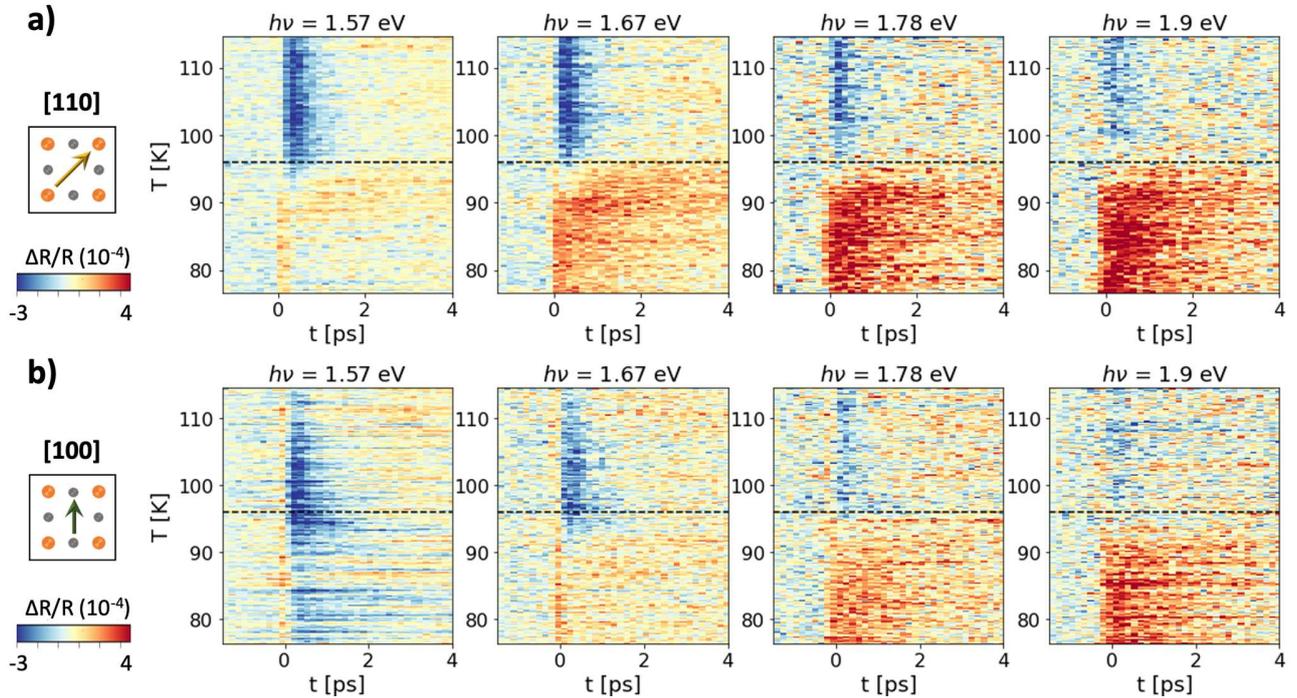

**Figure S5: Temperature-dependence upon sub-gap photoexcitation. a)** Transient reflectivity as function of temperature and pump-probe delay following the mid-IR pump pulse ($\phi_{MIR}$ = 28 µJ/cm$^2$) at different probe energies, indicated in the title of each panel. The pump pulse was polarized along [110], as illustrated by the sketch of the CuO$_2$ plaquette on the left. The black dashed line marks the sample critical temperature. **b)** Same as a) but for the pump polarized along [100]. What clearly emerges from the comparison of all the maps is that the critical temperature (marked by the dashed black line) cannot be univocally inferred from the measurements.

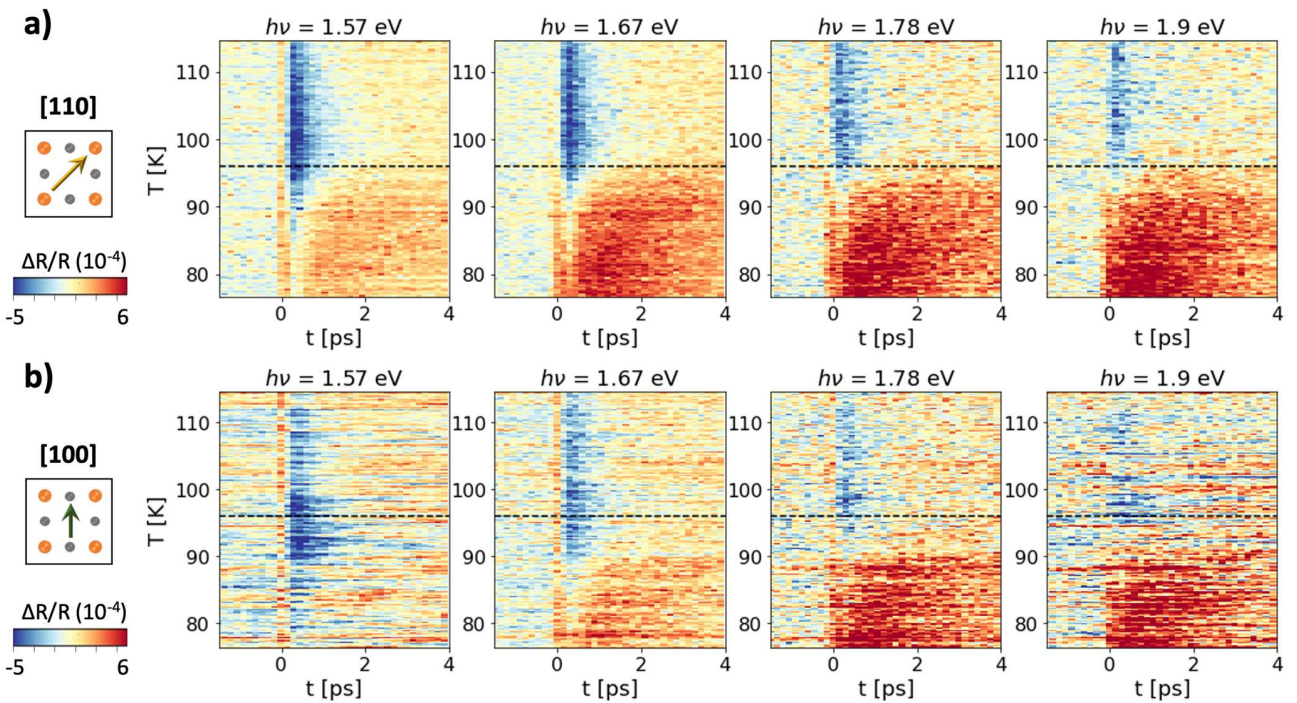

**Figure S6: Temperature-dependence upon above-gap photoexcitation. a-b)** Transient reflectivity as function of temperature and delay for different probe energies after photoexcitation by the near-IR pump ($\phi_{NIR}$ = 19 µJ/cm$^2$) polarized along [110] and [100].

# FIGURE S7. Differential fit of the temperature maps

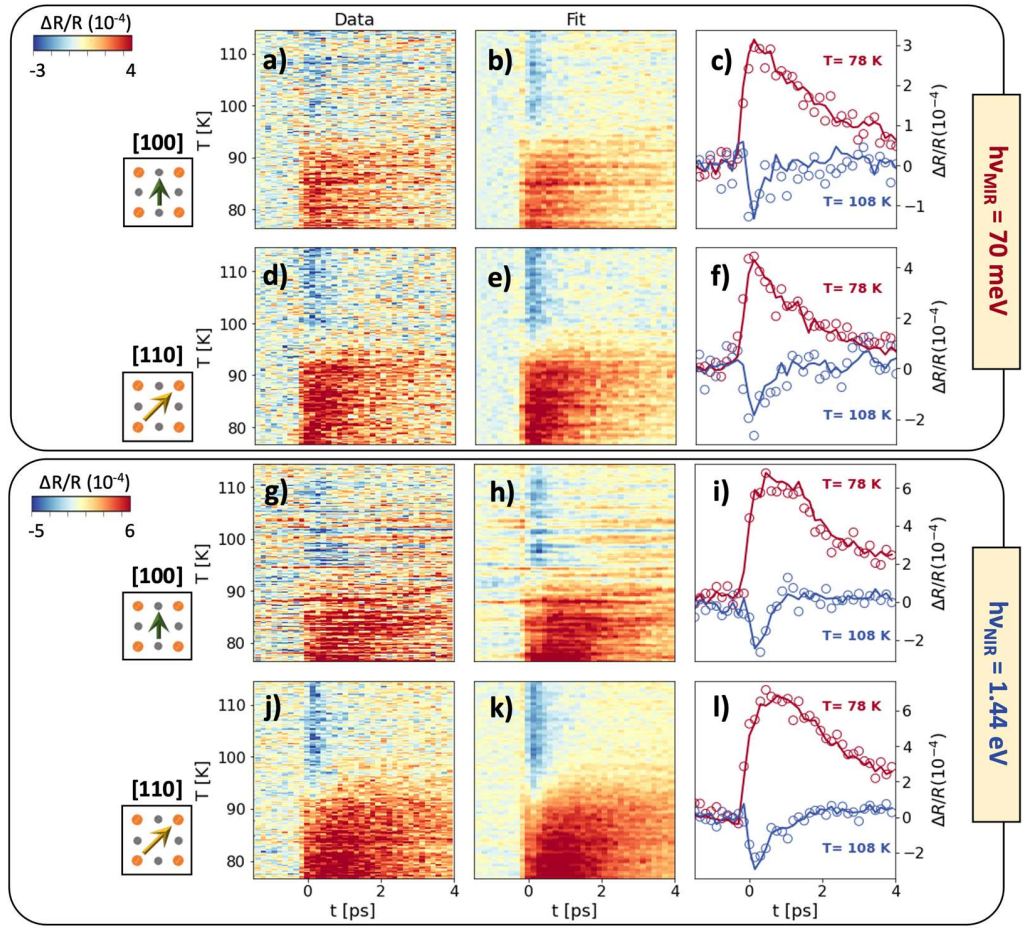

**Figure S7: a)** Temperature- and time-dependent reflectivity upon photoexcitation by the mid-IR (70 meV) pump polarized along the [100] measured at hν$_{probe}$=1.9 eV. **b)** Differential fit of a) obtained by the transient modification of the 2 eV oscillator. **c)** Temporal cuts at two distinct temperatures (SC and PG phase). The open circles are the data in a) and the solid line the fit in b). **d-e-f)** Same as a-b-c) but for the mid-IR pump polarization parallel to [110]. **g-l)** Same as a-f) but for the near-IR (1.44 eV) pump.

# Figure S8. Double-pump measurements in the superconducting phase

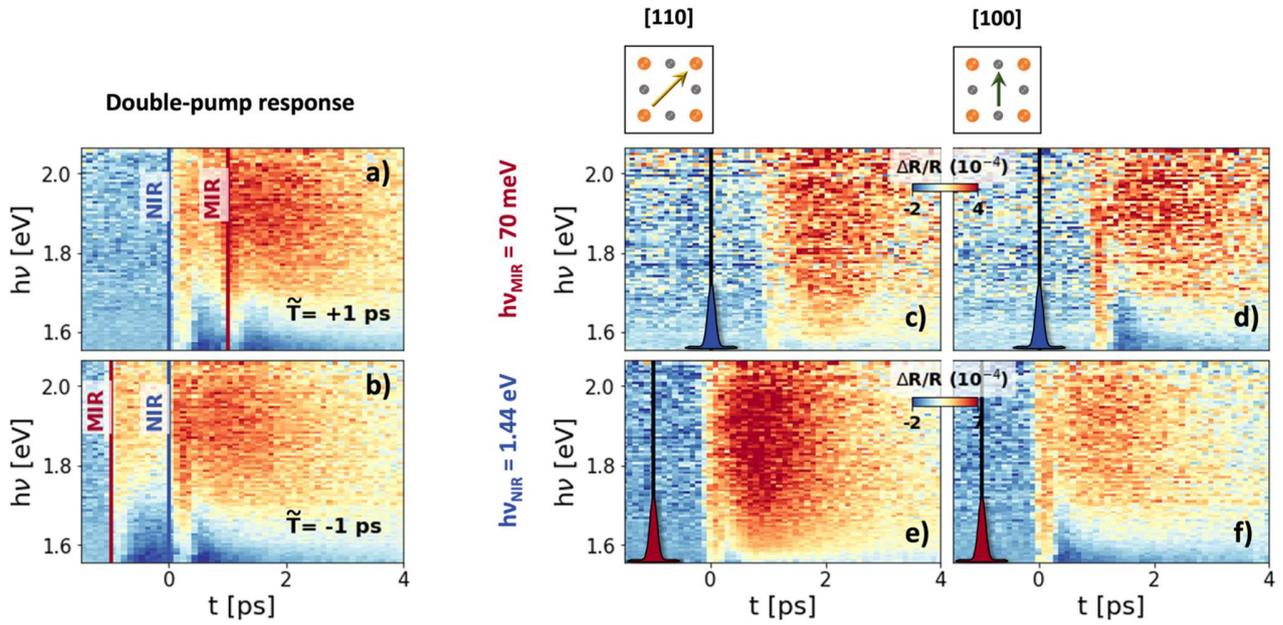

**Figure S8: a-b)** Example of the double-pump response measured in the experiment at T=74 K. The mid-IR pump impinges on the sample 1 ps after ($\tilde{T} = +1$ ps) and 1 ps before ($\tilde{T} = -1$ ps) the near-IR pump, respectively. The x-axis is the time delay between the near-IR pump and the white-light probe. **c)** Broadband transient reflectivity upon photoexcitation by the mid-IR pump on a system that has been previously (at t = 0 ps) excited by the above-gap pump, as indicated by the blue Gaussian-shaped pulse. Both the mid-IR and the near-IR polarizations are parallel to [110], as illustrated in the top sketch. The pump fluences are $\phi_{MIR} = 28$ μJ/cm$^2$ and $\phi_{NIR} = 19$ μJ/cm$^2$. **d)** Same as c) but for the pumps' polarization parallel to [100]. **e-f)** Same as c-d) but for a system that has been previously prepared by a mid-IR pump (red Gaussian at t = -1 ps) and then photoexcited by the above-gap pump.

# Figure S9. Spectral weight dynamics in the double-pump experiment ([110] axis)

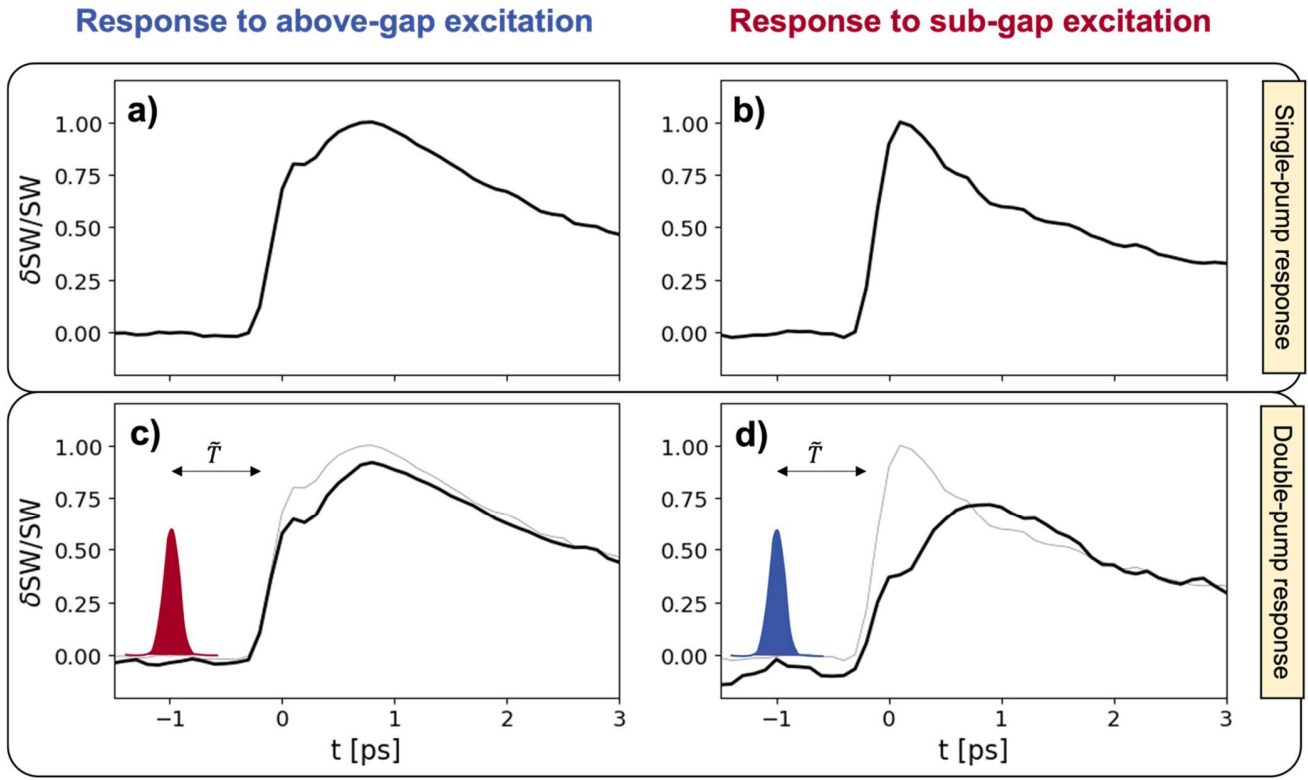

**Figure S9: a), b)** Normalized spectral weight dynamics extrapolated from the single-pump response to the above- and the sub-gap pulse, respectively. The polarization of the pumps is parallel to [110] and the sample temperature is 74 K. **c)** Spectral weight dynamics (black line) measured by the above-gap pulse when the system has been previously photoexcited by the mid-IR pulse at t=-1 ps (red Gaussian curve). **d)** Spectral weight dynamics (black line) measured by the sub-gap pulse in a sample that has previously (at t=-1 ps) interacted with the near-IR pulse. Grey curves in c) and d) are the single-pump spectral weight dynamics displayed in a) and b), over which the black curves have been renormalized.

# FIGURE S10. Central frequency dynamics in the double-pump experiment

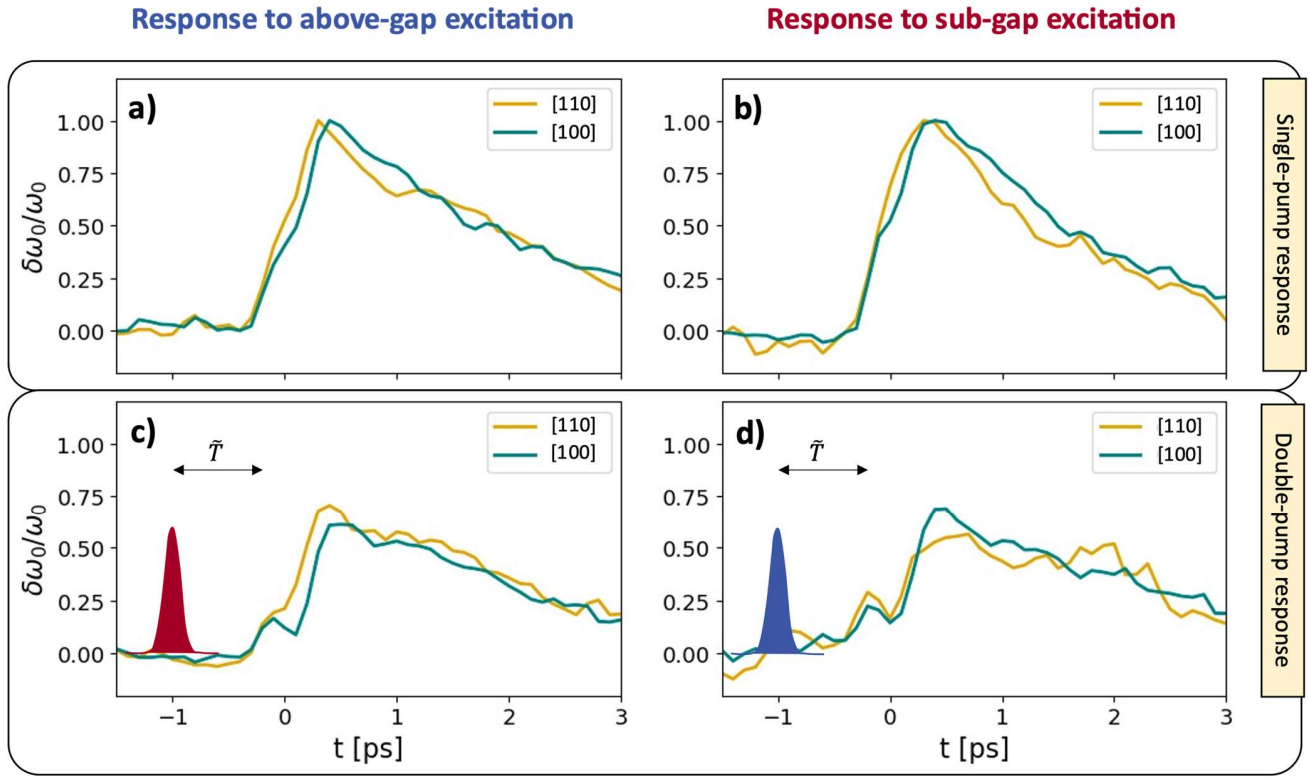

**Figure S10: a-b) a), b)** Normalized central frequency dynamics extrapolated from the single-pump response to the above- and the sub-gap pulse, respectively, for pumps polarized along [110] (gold) and [100] (teal). The sample temperature is 74 K. **c)** Central frequency dynamics measured by the above-gap pulse when the system has been previously photoexcited by the mid-IR pulse at t=-1 ps (red Gaussian curve). **d)** Central frequency dynamics measured by the sub-gap pulse in a sample that has previously (at t=-1 ps) interacted with the near-IR pulse.

# FIGURE S11. Double-pump dynamics

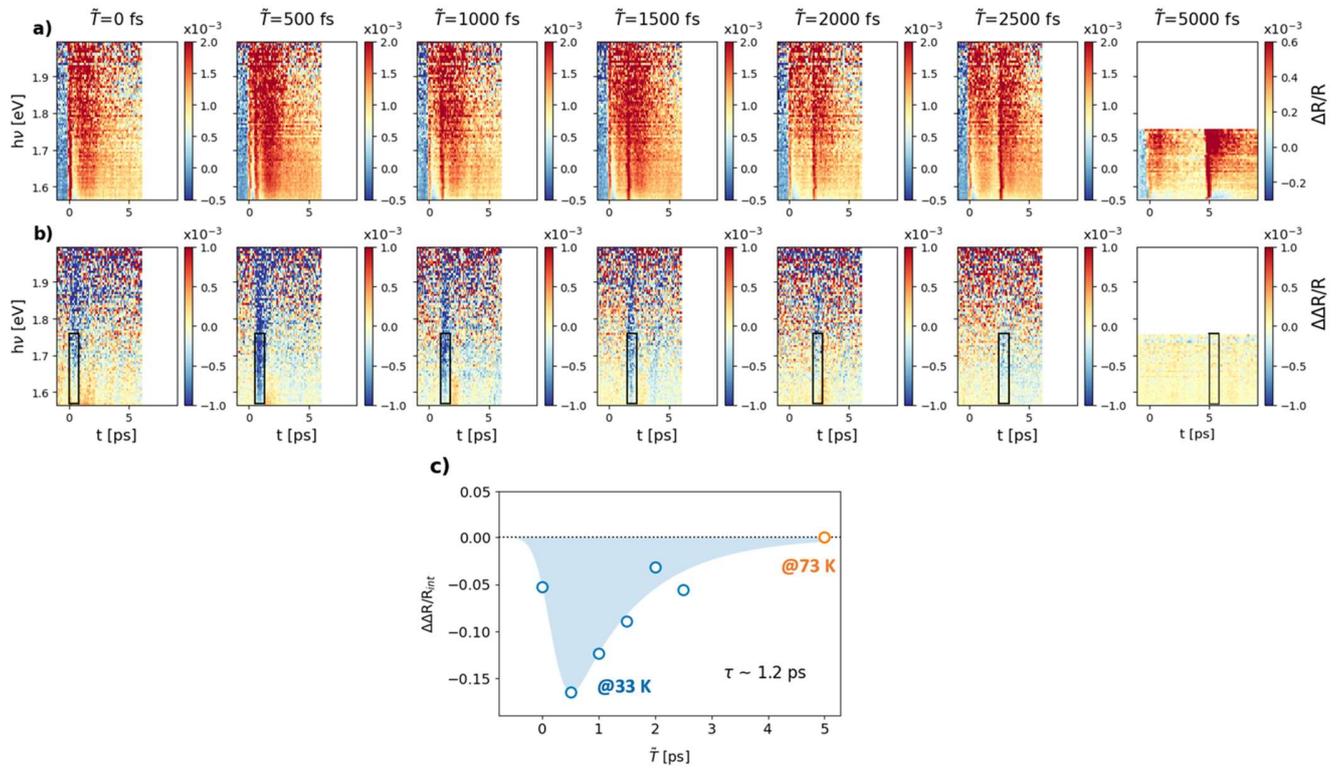

**Figure 11: a)** Double-pump response at different delays $\tilde{T}$ between the pumps: the near-IR pump impinges at t=0, while the mid-IR pump is delayed in time. Both pump polarizations are parallel to [110]. Sample temperature is 33 K, expect for the measurement at $\tilde{T}$=5000 fs where T=73 K. **b)** Same as a) but displaying the differential signal $\Delta\Delta R/R$, i.e., the double-pump response in a) after subtraction of the single-pump responses at every $\tilde{T}$. **c)** The datapoints indicate $\Delta\Delta R/R$ as function of $\tilde{T}$ integrated in the black area in b). The blue-shaded area is a fit to the data using a Gaussian convoluted with an exponential decay. The optimally-fitting decay time is τ~1.2 ps.